\documentclass[aps,showpacs,twocolumn,
superscriptaddress]{revtex4}
\usepackage{graphicx}
\usepackage{dcolumn}
\usepackage{bm}
\usepackage{color}
\usepackage[normalem]{ulem} 
\usepackage[dvipsnames]{xcolor} 
\usepackage{hyperref}
\hypersetup{
  colorlinks=true,        
  linkcolor=blue,         
  citecolor=cyan,         
}
\usepackage{mathrsfs}
\usepackage[utf8]{inputenc}
\usepackage{mathtools}
\usepackage{doi}
\usepackage{amsmath}
\usepackage{amssymb}

\newcommand{\sn}{\operatorname{\mathrm{sn}}}

\usepackage{scalerel}
\usepackage{tikz}
\usetikzlibrary{svg.path}

\definecolor{orcidlogocol}{HTML}{A6CE39}
\tikzset{
  orcidlogo/.pic={
    \fill[orcidlogocol] svg{M256,128c0,70.7-57.3,128-128,128C57.3,256,0,198.7,0,128C0,57.3,57.3,0,128,0C198.7,0,256,57.3,256,128z};
    \fill[white] svg{M86.3,186.2H70.9V79.1h15.4v48.4V186.2z}
                 svg{M108.9,79.1h41.6c39.6,0,57,28.3,57,53.6c0,27.5-21.5,53.6-56.8,53.6h-41.8V79.1z M124.3,172.4h24.5c34.9,0,42.9-26.5,42.9-39.7c0-21.5-13.7-39.7-43.7-39.7h-23.7V172.4z}
                 svg{M88.7,56.8c0,5.5-4.5,10.1-10.1,10.1c-5.6,0-10.1-4.6-10.1-10.1c0-5.6,4.5-10.1,10.1-10.1C84.2,46.7,88.7,51.3,88.7,56.8z};
  }
}

\newcommand\orcidicon[1]{\href{https://orcid.org/#1}{\mbox{\scalerel*{
\begin{tikzpicture}[yscale=-1,transform shape]
\pic{orcidlogo};
\end{tikzpicture}
}{|}}}}

\begin{document}

\title{
Orbits of particles with magnetic dipole moment around magnetized Schwarzschild black holes: Applications to S2 star orbit}

%
\author{Uktamjon Uktamov} \email{uktam.uktamov11@gmail.com}
\affiliation{Institute of Fundamental and Applied Research, National Research University TIIAME, Kori Niyoziy 39, Tashkent 100000, Uzbekistan}

\author{Mohsen Fathi\orcidicon{0000-0002-1602-0722}}
\email{mohsen.fathi@ucentral.cl}
\affiliation{Vicerrector\'{i}a Acad\'{e}mica, Universidad Central de Chile,  Toesca 1783, Santiago 8320000, Chile}
\affiliation{Facultad de Ingenier\'{i}a y Arquitectura, Universidad Central de Chile,  Av. Santa Isabel 1186, 8330563, Santiago, Chile}
 
\author{Javlon Rayimbaev}
\email{javlon@astrin.uz}
\affiliation{Institute of Fundamental and Applied Research, National Research University TIIAME, Kori Niyoziy 39, Tashkent 100000, Uzbekistan}
\affiliation{University of Tashkent for Applied Sciences, Str. Gavhar 1, Tashkent 100149, Uzbekistan}
\affiliation{Urgench State University, Kh. Alimjan str. 14, Urgench 221100, Uzbekistan}

\author{Ahmadjon~Abdujabbarov} \email{ahmadjon@astrin.uz} 
\affiliation{Ulugh Beg Astronomical Institute, Astronomy St 33, Tashkent 100052, Uzbekistan}
\affiliation {Shahrisabz State Pedagogical Institute, Shahrisabz Str. 10, Shahrisabz 181301, Uzbekistan}
\affiliation{Tashkent State Technical University, Tashkent 100095, Uzbekistan}



%

\date{\today}

\begin{abstract}
This study provides a comprehensive analytical investigation of the bound and unbound motion of magnetized particles orbiting a Schwarzschild black hole immersed in an external asymptotically uniform magnetic field, which includes all conceivable types of bounded and unbounded orbits. In particular, for planetary orbits, we perform a comparative analysis of our findings with the observed position of the S2 star carrying magnetic dipole moment around Sagittarius A* (Sgr A*). We found maximum and minimum values for the parameter of magnetic interaction between the magnetic dipole of the star and the external magnetic field, as well as the energy and angular momentum of the S2 star. As a result, we obtain estimations of the magnetic dipole of the star in order of $10^6 \rm \ G\cdot cm^{3}$. Additionally, we explore deflecting trajectories akin to gravitational Rutherford scattering. In obtaining the solutions for the orbital equations, we articulate the elliptic integrals and Jacobi elliptic functions, and illustrative figures and simulations augment our study.
\end{abstract}

\maketitle

\section{Introduction}



The investigation of free-falling objects in gravitational fields links geometry and gravity through the general theory of relativity, where the curvature of spacetime plays the principal role. This fundamental understanding gained popularity early once it was recognized that planets and light follow geodesics. The outstanding discovery, right after the Schwarzschild solution proposition \cite{1916SPAW189S}, was the verification of light deflection during the 1919 solar eclipse expedition \cite{1920RSPTA.220..291D}. The precise evaluation of the anomalous precession in the perihelion of Mercury \cite{RevModPhys.19.361}, which constitute the primary standard tests of general relativity. However, the complex nature of the partial differential equations governing particle dynamics in curved spacetimes necessitates simplified or numerical solutions to compare theoretical results with observational tests. 

Analytical expressions have several advantages since they serve as benchmarks for numerical methods and, hence, they facilitate comprehensive parameter space exploration for predicting astrophysical observables. Efforts to find exact analytical solutions for geodesic equations of particles have intensified since Hagihara's studies in 1931 \cite{1930JaJAG...8...67H}, followed by works by Darwin, Mielnik, and Pleba\'{n}ski \cite{noauthor_gravity_1959,noauthor_gravity_1961,Mielnik:1962}. These methods, which are based on the theories of elliptic functions and modular forms, were studied by nominated nineteenth-century mathematicians such as Jacobi \cite{jacobi_2013}, Abel \cite{abel_2012}, Riemann \cite{Riemann:1857,Riemann+1866+161+172}, and Weierstrass \cite{Weierstrass+1854+289+306} (see also Ref. \cite{baker_abelian_1995} for a complete textbook review on these discoveries). Particularly noteworthy is the application of modular forms in solving (hyper-)elliptic integrals involved in particle trajectory studies, garnering significant attention in recent decades. These methods, rooted in the theories of elliptic functions and modular forms, have been the focus of numerous investigations analyzing trajectory curves in black hole spacetimes inferred from general relativity and its extensions (see, for example, Refs. \cite{kraniotis_general_2002, kraniotis_compact_2003, kraniotis_precise_2004, kraniotis_frame_2005, cruz_geodesic_2005, kraniotis_periapsis_2007, hackmann_complete_2008, hackmann_geodesic_2008, hackmann_analytic_2009, hackmann_complete_2010, olivares_motion_2011, kraniotis_precise_2011, cruz_geodesic_2013, villanueva_photons_2013, kraniotis_gravitational_2014, soroushfar_analytical_2015, villanueva_gravitational_2015,soroushfar_detailed_2016, hoseini_analytic_2016, hoseini_study_2017, fathi_motion_2020, fathi_classical_2020, fathi_gravitational_2021, gonzalez_null_2021, fathi_analytical_2021, kraniotis_gravitational_2021, fathi_study_2022, soroushfar_analytical_2022, battista_geodesic_2022, fathi_spherical_2023, FATHI2023169401}).

Building upon this scientific foundation, contemporary research continues to explore the properties and dynamics of particles falling in gravitational fields in diverse astrophysical contexts. bound and unbound particle orbits, in particular, represent fundamental aspects of motion that offer valuable insights into the structure and evolution of the universe. Bound orbits, characterized by closed trajectories around central masses, are found extensively in celestial systems, ranging from planetary orbits in the solar system to the motion of stars in galaxies \cite{perek_galactic_1962,barbieri_galactic_2002}. These orbits contribute to astrophysical objects' stability and long-term dynamics, and provide essential constraints on theoretical models of gravitational dynamics and cosmological evolution.

On the other hand, unbound particle orbits extend indefinitely through space, either escaping to infinity (deflecting trajectories) or falling into gravitational singularities (plunge or captured orbits). These orbits play a critical role in understanding cosmic phenomena such as the formation and evolution of galaxies and the dynamics of stellar populations within galactic clusters (see, for example, Ref. \cite{zhong_exploring_2023} and the book \cite{binney_galactic_2011}). They are also useful for modeling particle orbits within dark matter halos \cite{behroozi_unbound_2013}. Unbound orbits provide valuable information about the escape velocity of celestial bodies, shedding light on the mechanisms underlying stellar ejection and galactic interactions. Moreover, the study of unbound orbits offers insights into the gravitational interactions between celestial objects and their surrounding environments, contributing to our understanding of the large-scale structure and evolution of the universe \cite{binney_galactic_2011}.

The study of trajectories of magnetized particles around magnetized astrophysical objects, particularly black holes, is of great importance in understanding the nature of astrophysical phenomena. Magnetized black holes, characterized by intense magnetic fields generated by rapidly rotating neutron stars or stellar-mass black holes, exhibit intricate interactions between gravitational and electromagnetic forces that give rise to a diverse range of astrophysical phenomena \cite{aliev_magnetized_1989,li_observational_2002,meier_black_2012,kruglov_magnetized_2017,ghosh_astrophysical_2021,kopacek_magnetized_2023,karas_magnetized_2023,Khan:2024jez}. Charged particles, such as electrons and protons, spiraling along magnetic field lines near these objects undergo complex dynamics, including acceleration, deflection, and emission of radiation. These processes play a crucial role in shaping the observational properties of various astrophysical sources, including pulsars, gamma-ray bursts, and active galactic nuclei \cite{Chang_2022,Meszaros_2019,Novikov1986}. By studying the trajectories of magnetized particles, researchers can gain insights into the mechanisms driving particle acceleration, the formation of relativistic jets, and the emission of high-energy radiation observed across the electromagnetic spectrum. Furthermore, understanding the interplay between gravitational and electromagnetic forces near magnetized astrophysical objects offers valuable insights into fundamental aspects of plasma physics, high-energy astrophysics, and general relativity \cite{nishikawa_general_2005,moscibrodzka_general_2016,mizuno_grmhd_2022}. 

Given the importance of this subject, our study aims to investigate the dynamics of magnetized particles affected by the magnetic field of an astrophysical object, in this case, a magnetized Schwarzschild black hole. In this study, we analyze the dynamics and motion of particles possessing magnetic dipole moments analytically as they move in the exterior geometry of a Schwarzschild black hole with an associated magnetic field. We utilize elliptic integrals to express exact analytical solutions to the equations of motion governing the behavior of these particles. Particularly, the deflecting trajectories are akin to a type of gravitational Rutherford scattering, which is also treated analytically.

To accomplish our objective, we structure this paper as follows: In Section \ref{sec:Sch_mag}, we introduce the spacetime under study and develop a Lagrangian formalism, leading to the equations of motion for the test particles. In Section \ref{sec:bound}, we initiate our main study by providing a general classification of the possible orbits based on the radial profile of the effective potential experienced by the particles. In this section, we provide a precise derivation of the planetary orbits with different eccentricities, also assessed by observational tests regarding the S-star orbits at the center of the Milky Way. In Section \ref{sec:unbound}, we continue our study by investigating the unbound orbits consisting of deflecting trajectories. These trajectories are deflected beyond the black hole or inevitably fall into the event horizon. The former is termed gravitational Rutherford scattering in Refs. \cite{villanueva_gravitational_2015,fathi_gravitational_2021}. To complete this investigation, in Section \ref{sec:radial}, we present a fully analytical study of the purely radial trajectories. We conclude in Section \ref{sec:conclusion}. Throughout the paper, we work in the geometrized system of units where $G=c=1$, we adopt the sign convention $(- + +\, +)$, and wherever primes appear, they denote differentiations for the radial coordinate.

\section{Motion of magnetized particles around a Schwarzschild black hole immersed in magnetic fields
}\label{sec:Sch_mag}

In this section, we derive the equations of motion governing magnetized test particles orbiting a Schwarzschild black hole in the presence of an externally applied asymptotically uniform magnetic field. The line element describing the exterior spacetime of a Schwarzschild black hole with mass $M$ is given by
\begin{equation}\label{1}
ds^2=-f(r)dt^2+f(r)^{-1}dr^2+r^2d\Omega^2,
\end{equation}
in the usual Schwarzschild coordinates $x^\alpha=(t,r,\theta,\phi)$, with $d\Omega^2=d\theta^2+\sin^2\theta d\phi^2$, where the lapse function is given by
\begin{equation}
  f(r)=1-\frac{2M}{r}.  
  \label{eq:lapse0}
\end{equation}

\subsection{Magnetization of Schwarzschild-like black holes}

Here, we assume that the black hole is surrounded by an external, asymptotically uniform magnetic field with a constant value of $B$ at infinity. According to Wald's approach, the four-potential, being solutions of Maxwell's equations, can be expressed as $A^\mu=\frac{B}{2}\xi^\mu_\phi$ \cite{wald_general_1984}, where $\xi_\phi^\mu=(0,0,0,1)$ represents a space-like Killing vector. Consequently, we can immediately obtain the non-vanishing components of the electromagnetic tensor as
\begin{subequations}\label{3}
\begin{align}
 & F_{r\phi}=Br\sin^2{\theta}, \\
 & F_{\theta\phi}=Br^2\sin{\theta}\cos\theta.  
\end{align}
\end{subequations}
The components of the magnetic field around the black hole, as measured by zero angular momentum observers (ZAMOs), can be calculated as
\begin{eqnarray}\label{fields}
B^i &=& \frac{1}{2}\eta^{i \beta \sigma \mu} F_{\beta \sigma} w_{\mu},
\end{eqnarray}
where $i=(r,\theta,\phi)$, $\bm{w}$ represents the four-velocity of the ZAMOs with $w^{\mu}=1/\sqrt{f(r)}(1,0,0,0)$, and $\eta_{\alpha \beta \sigma \gamma}$ denotes the pseudo-tensorial form of the Levi-Civita symbol $\epsilon_{\alpha \beta \sigma \gamma}$ with the following relations: 
\begin{eqnarray}
&&\eta_{\alpha \beta \sigma \gamma}=\sqrt{-g}\,\epsilon_{\alpha \beta \sigma \gamma},\quad \eta^{\alpha \beta \sigma \gamma}=-\frac{1}{\sqrt{-g}}\epsilon^{\alpha \beta \sigma \gamma},
\end{eqnarray}
where $g={{\rm{det}}|g_{\mu \nu}|}=-r^4\sin^2\theta$, corresponding to the spacetime metric (\ref{1}). 
Hence, one can obtain expressions for the non-zero orthonormal components of the magnetic field using the tetrads carried by the ZAMOs \cite{Rezzolla01c}. Furthermore, the non-zero orthonormal components of the magnetic field measured by the ZAMOs are calculated as
\begin{subequations}\label{4}
\begin{align}
    & B^{\Hat{\theta}}=\sqrt{f(r)}\,B\sin{\theta},\\
    & B^{\Hat{r}}=B\cos{\theta}.
\end{align}
\end{subequations}

\subsection{Magnetized particle dynamics}

The dynamics of polarized particles within the spacetime described by the line element \eqref{1} can be expressed using the Lagrangian \cite{2004PhRvD..70b4012P}
\begin{eqnarray}\label{5}
    \mathscr{L}=\frac{1}{2}(m+U)g_{\mu\nu}u^\mu u^\nu-\frac{1}{2}\kappa U,
\end{eqnarray}
where $\kappa=1$ and the magnetic interaction term is represented by $U=D^{\mu\nu}F_{\mu\nu}$. The calculation of the magnetic interaction term has been undertaken by various authors (see, for example, Ref. \cite{2003CQGra..20..469D}). Specifically, it is given by $U=\mu_{\Hat{\alpha}}B^{\Hat{\alpha}}$. Furthermore, suppose we limit the motion of the magnetized particles to the equatorial plane and assume that the magnetic dipole of the particles is perpendicular to this plane. In that case, we can express the components of the magnetic moment four-vectors as $\mu^{\hat{\alpha}}=(0,0,\mu,0)$. Consequently, we can derive the conjugate momenta as 
\begin{eqnarray}\label{6}
  p_\mu=\frac{\partial\mathscr{L}}{\partial \Dot{x}^\mu}=(m+U)g_{\mu\nu}u^\nu,  
\end{eqnarray}
based on which, the two constants of motion 
\begin{eqnarray}\label{7}
    -\mathcal{E}&=&\bigr[1+\beta\mathcal{F}(r,\theta)\bigr]g_{tt}\Dot{t},\label{7a}
\\ 
    l&=&\bigr[1+\beta\mathcal{F}(r,\theta)\bigr]g_{\phi\phi}\Dot{\phi},\label{7b}
\end{eqnarray}
can be defined in accordance with the symmetries of the spacetime, where $\beta=\mu B/m$, and $\mathcal{F}(r,\theta)=\sqrt{f(r)}\sin{\theta}$. Here, $\mathcal{E}=E/m$ and $l=L/m$ represent the magnetized particle's specific energy and specific angular momentum, respectively. Furthermore, one can express the Hamilton-Jacobi equation while accounting for the magnetic field as follows:
\begin{eqnarray}\label{8}
   g^{\mu\nu}\frac{\partial {\cal S}}{\partial x^\mu}\frac{\partial {\cal S}}{\partial x^\nu}=-m^2\left(1-\frac{U}{m}\right)^2. 
\end{eqnarray}
with $\mathcal{S}$ denoting the Jacobi action, and motion restricted to the equatorial plane, the action can be expressed as ${\cal S}=-Et+L\phi+{\cal S}_r(r)$, and in this context. Therefore, the radial motion can be easily described in terms of the following equation:
\begin{eqnarray}\label{9}
    \dot{r}^2=\mathcal{E}^2-V_{\mathrm{eff}},
\end{eqnarray}
in which
\begin{eqnarray}\label{10}
V_{\mathrm{eff}}(r;l)=f(r)\left(\left[1-{\beta\mathcal{F}(r,\pi/2)}\right]^2+\frac{l^2}{r^2}\right),
\end{eqnarray}
 do the test particles feel the effective potential, {which takes the following form in the weak electromagnetic interaction limit (at $\beta \ll 1$), considering the linear order of $\beta$:
\begin{eqnarray}\label{10 linerized}
V_{\mathrm{eff}}(r;l)=f(r)\left(1-2{\beta\mathcal{F}(r,\pi/2)}+\frac{l^2}{r^2}\right).
\end{eqnarray}}
 \begin{figure*}[ht!]
\includegraphics[width=0.324\textwidth]{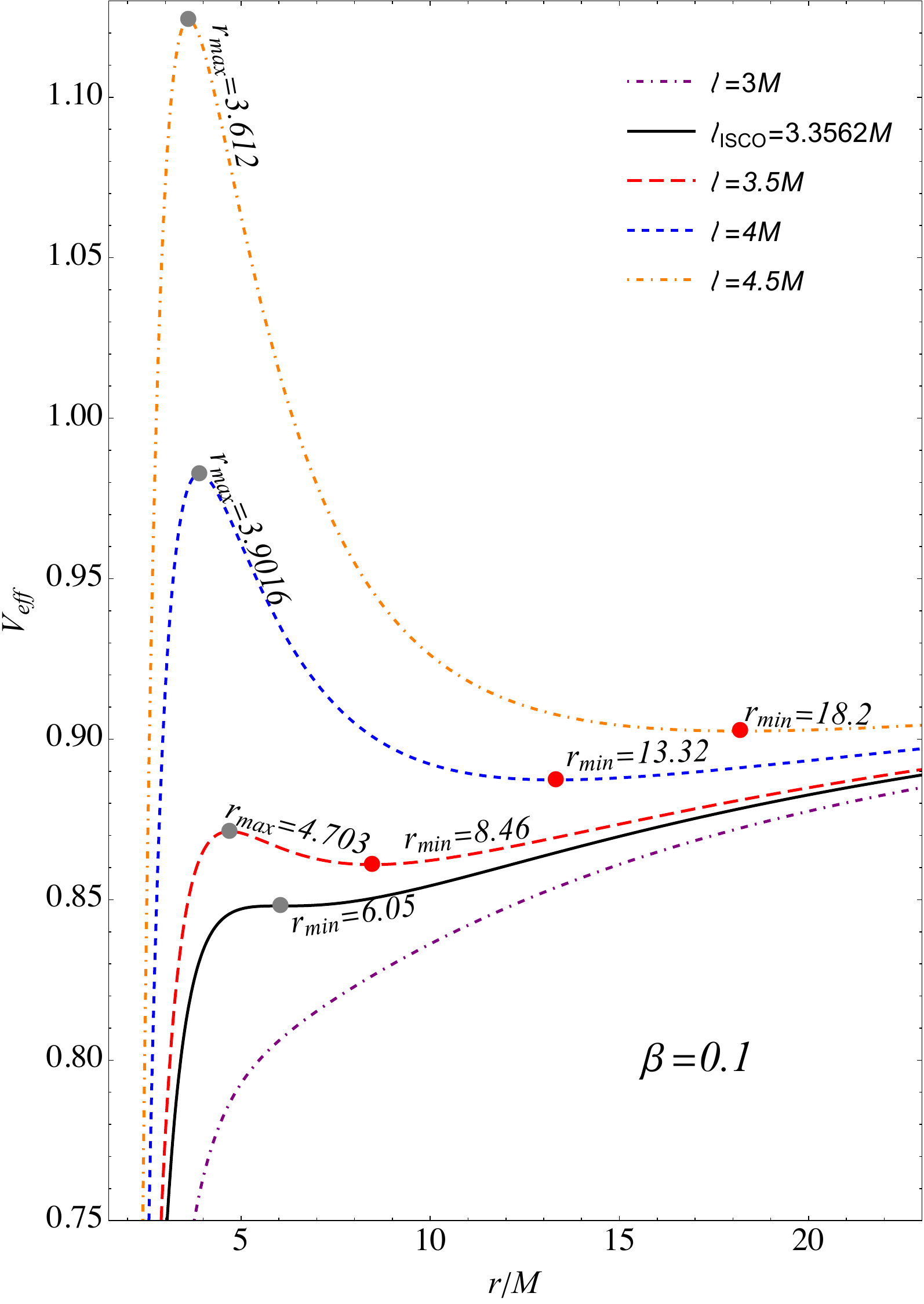}
\includegraphics[width=0.324\textwidth]{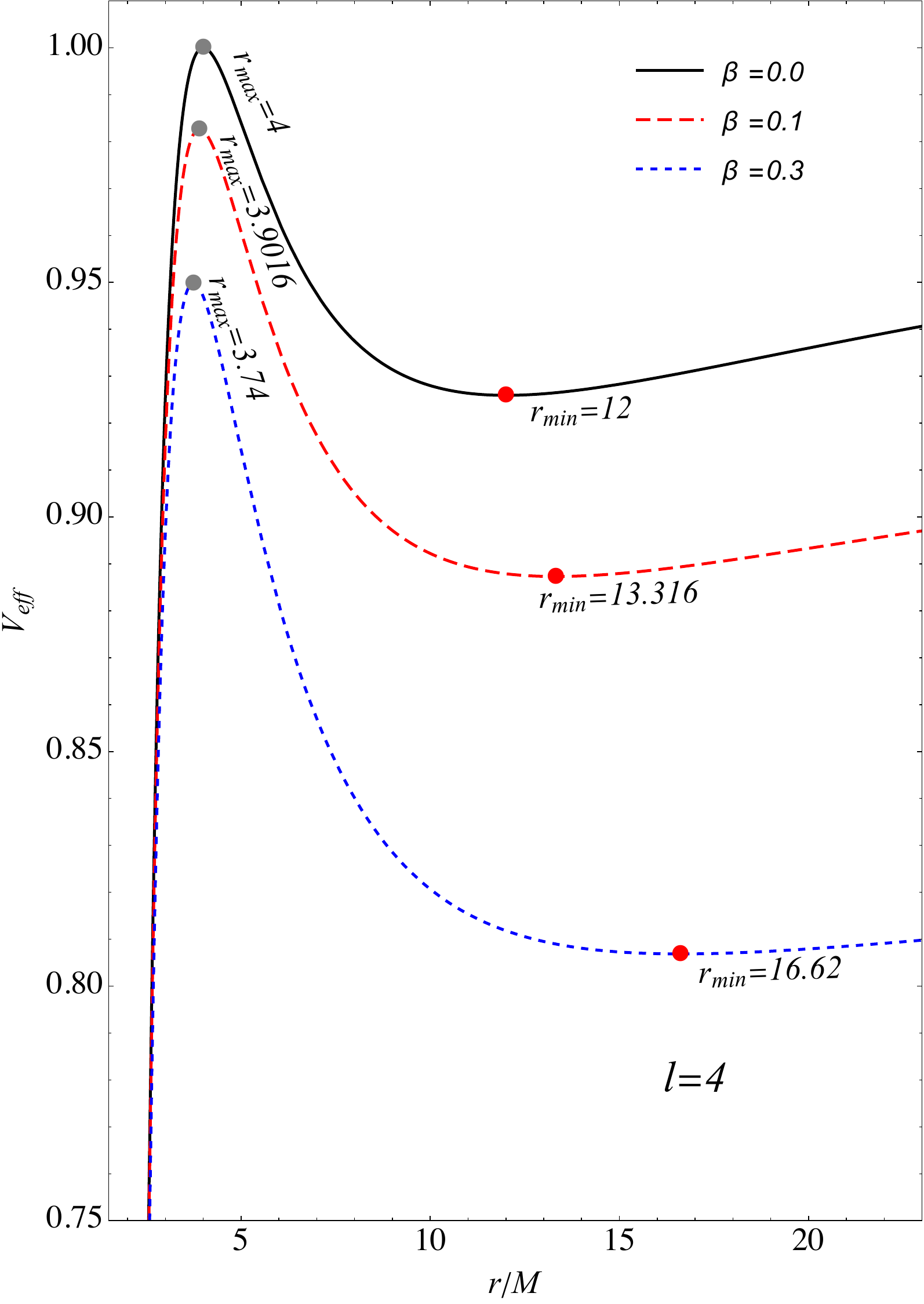}
\includegraphics[width=0.327\textwidth]{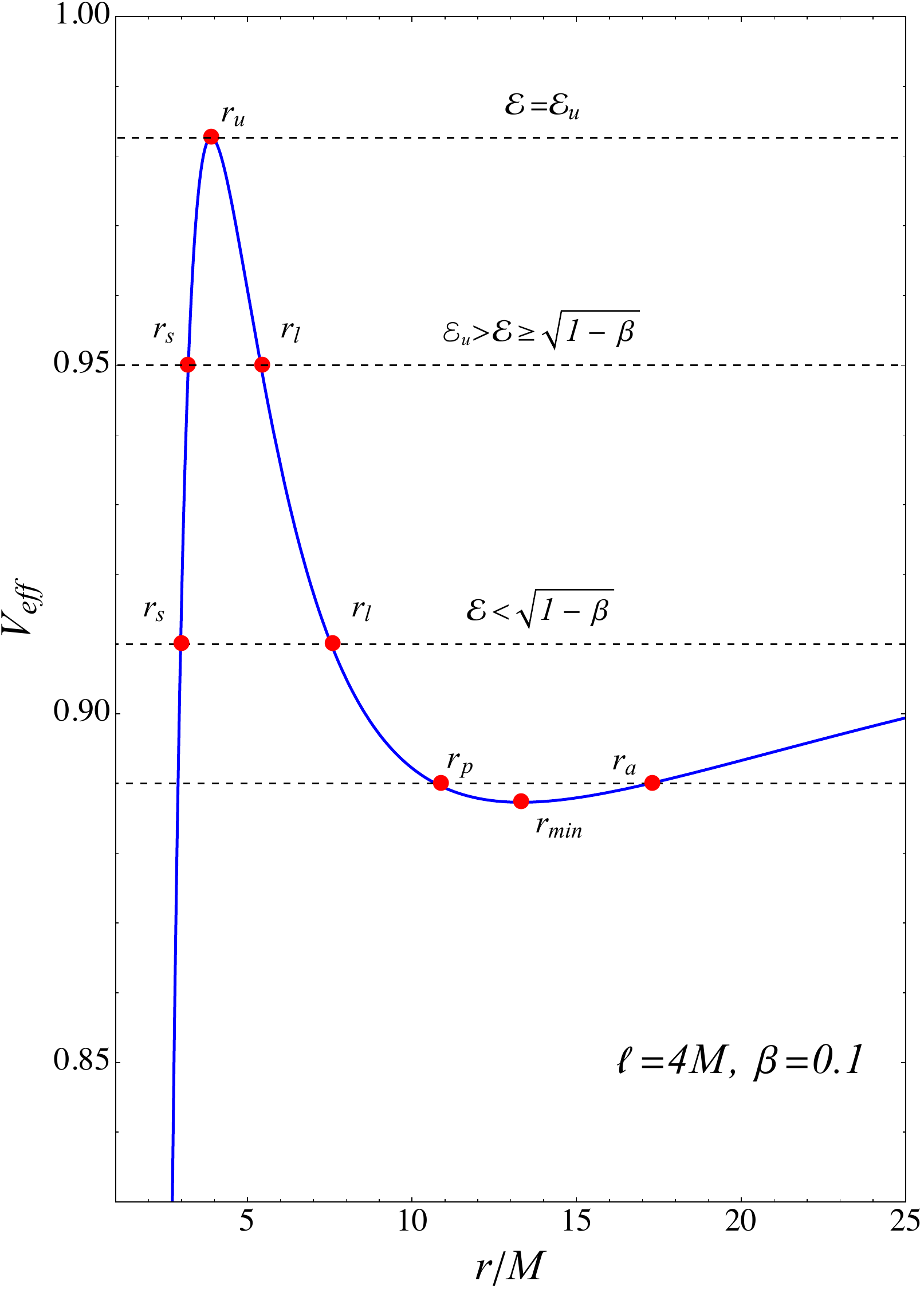}
\caption{The radial profile of the effective potential is depicted for fixed values of the magnetic coupling parameter $\beta$ (left panel), and the specific angular momentum $l$ (middle panel). The minima ($r_{\text{min}}$) in the effective potential corresponded to stable circular orbits, while the maxima ($r_{\text{max}}$) correspond to unstable circular orbits. Additionally, the turning points $r_s$, $r_l$, $r_p$, $r_a$, and the radius of the unstable circular orbit $r_u$ are indicated in the right panel.
\label{effective}}
\end{figure*}
In Fig. \ref{effective}, we illustrate how the radial dependence of the effective potential varies with the specific angular momentum (left panel) and the magnetic interaction parameter $\beta$ (middle panel). The maximum of the effective potential denotes instability. Furthermore, these plots indicate that increasing the specific angular momentum $l$ of the magnetized particles and the magnetic interaction parameter $\beta$ leads to an increase in the stable circular orbits (at $r_{\text{min}}$) of the test particles with magnetic dipole momentum. Conversely, an increase in the specific angular momentum of the magnetized particle results in a decrease in the unstable circular orbits (at $r_{\text{max}}$) of the particles with magnetic dipole moment. Moreover, escalating the magnetic coupling parameter $\beta$ increases the unstable circular orbits of the magnetized particles. Additionally, Fig. \ref{effective} presents the radial profile of the effective potential with the specific angular momentum $l_{\text{ISCO}}$ of the magnetized particles in the Innermost Stable Circular Orbit (ISCO) for a given value of $\beta=0.1$. Consequently, Fig. \ref{effective} indicates that for specific angular momentum values $l>l_{\text{ISCO}}$, there are two extreme points in the effective potential, namely $r_{\text{max}}$ corresponding to unstable circular orbits and $r_{\text{min}}$ corresponding to stable circular orbits. For the specific angular momentum value $l=l_{\text{ISCO}}$, there is only one extreme point $r_{\text{min}}$, corresponding to the stable circular orbit. However, for specific angular momentum values $l<l_{\text{ISCO}}$, there are no extreme points in the effective potential, indicating the absence of any circular orbits in this case. Additionally, insights into the orbits of magnetized test particles can be obtained from the effective potential. In the right panel of Fig. \ref{effective}, we depict the turning points $r_s$ and $r_l$ corresponding to the deflecting trajectories. Furthermore, concerning the planetary bound orbits, we have also denoted the turning points $r_p$ (periapsis), the smallest orbital separation, and $r_a$ (apoapsis), the most significant orbital separation, of particles possessing magnetic dipole moment orbiting a Schwarzschild Black Hole within magnetized fields. We will later explore that test particles establish bound orbits if their specific energy $\mathcal{E}$ is less than $\sqrt{1-\beta}$ and unbound orbits if $\mathcal{E}_u > \mathcal{E} \geq \sqrt{1-\beta}$, where $\mathcal{E}_u^2=V_{\text{eff}}(r_u)$ represents the energy corresponding to the unstable circular orbits at $r_u=r_{\text{max}}$.

One can utilize Eqs. (\ref{7}) and (\ref{9}) to derive the general equation of motion in the equatorial plane for angular trajectories. This yields the differential equation
\begin{multline}\label{11}
    \left(\frac{dr}{d\phi}\right)^2=\frac{r^4\left(1+\beta \sqrt{f(r)}\right)^2}{l^2}\\ 
    \times\left[\mathcal{E}^2-f(r)\left(1+\frac{l^2}{r^2}-\beta\sqrt{f(r)}\right)\right].
\end{multline}
{By simplifying Eq. \eqref{10} to the first order in $\beta$, we obtain
\begin{multline}\label{11 linerized}
    \left(\frac{dr}{d\phi}\right)^2 = \frac{r^4}{l^2} \Biggl\{\mathcal{E}^2 \left(1+2 \beta  \sqrt{f(r)}\right)\\ 
    -f(r) \left[\beta  \sqrt{f(r)} \left(1+\frac{2 l^2}{r^2}\right)+1+\frac{l^2}{r^2}\right]\Biggr\}.
\end{multline}}
In the forthcoming paper sections, we will employ this equation to analyze the bound and unbound orbits of magnetized test particles in the spacetime under consideration.

\section{The bound orbits}\label{sec:bound}

\begin{figure*}[t]
\includegraphics[width=0.3245\textwidth]{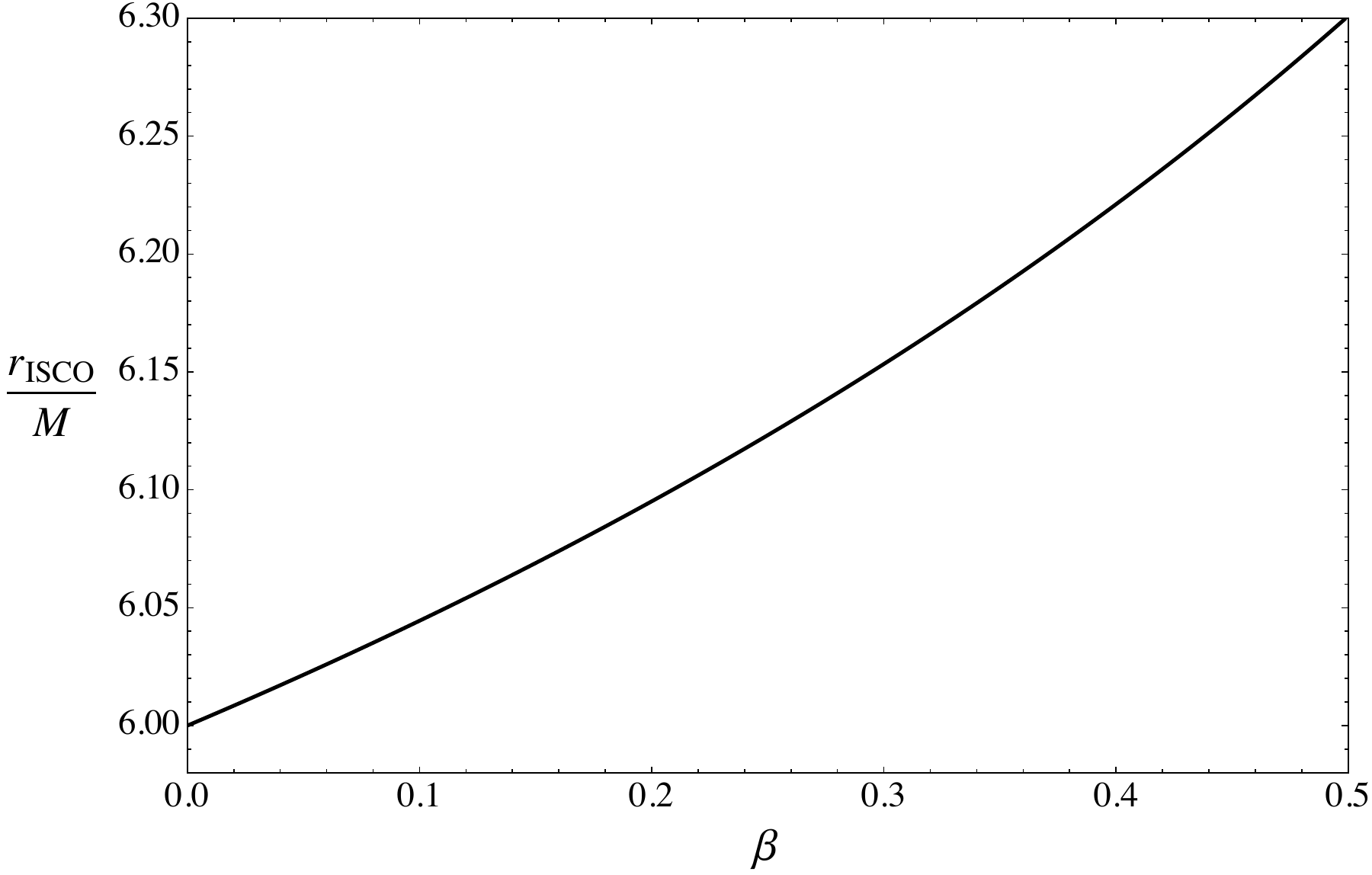}
\includegraphics[width=0.3245\textwidth]{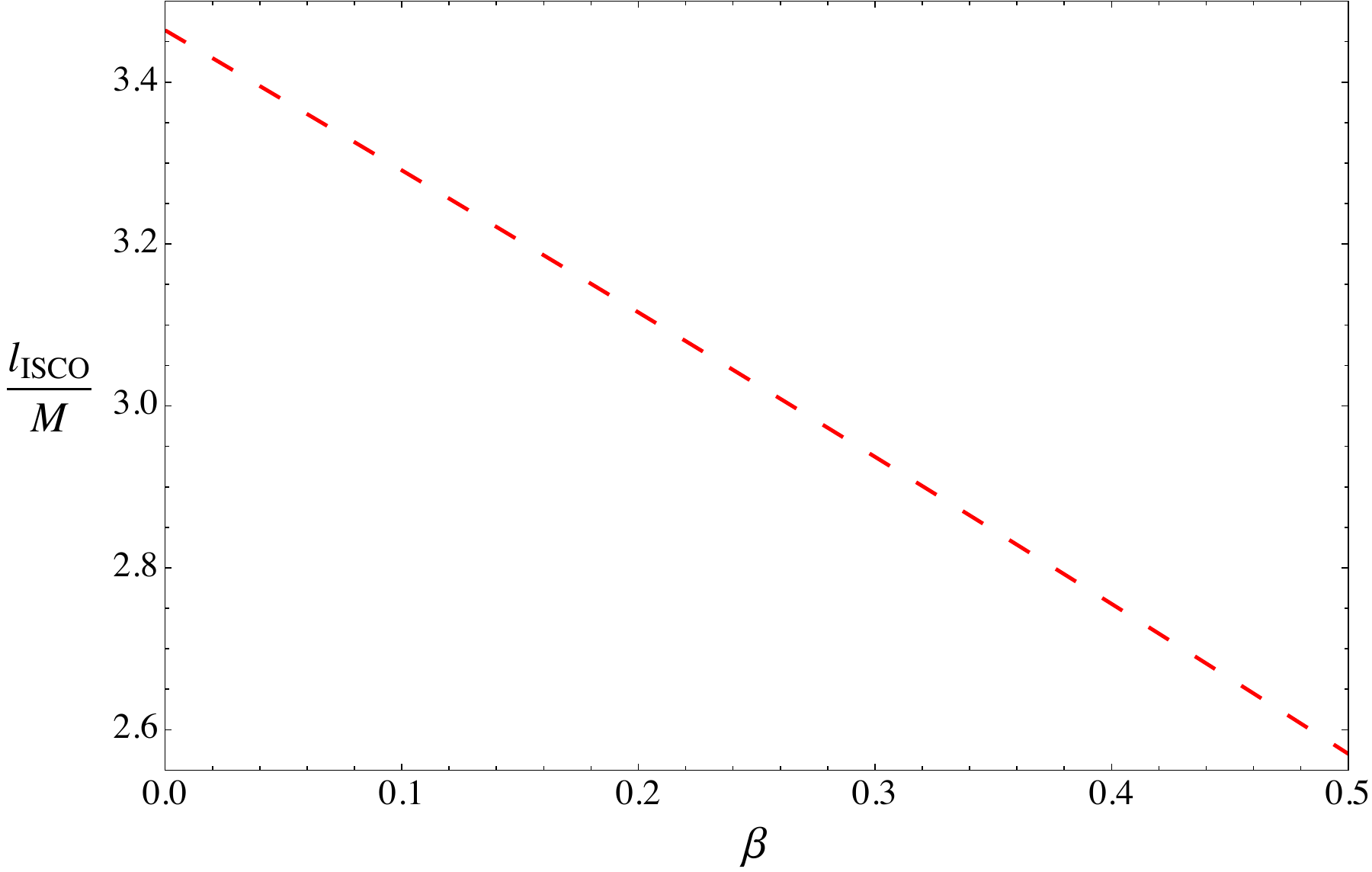}
\includegraphics[width=0.3245\textwidth]{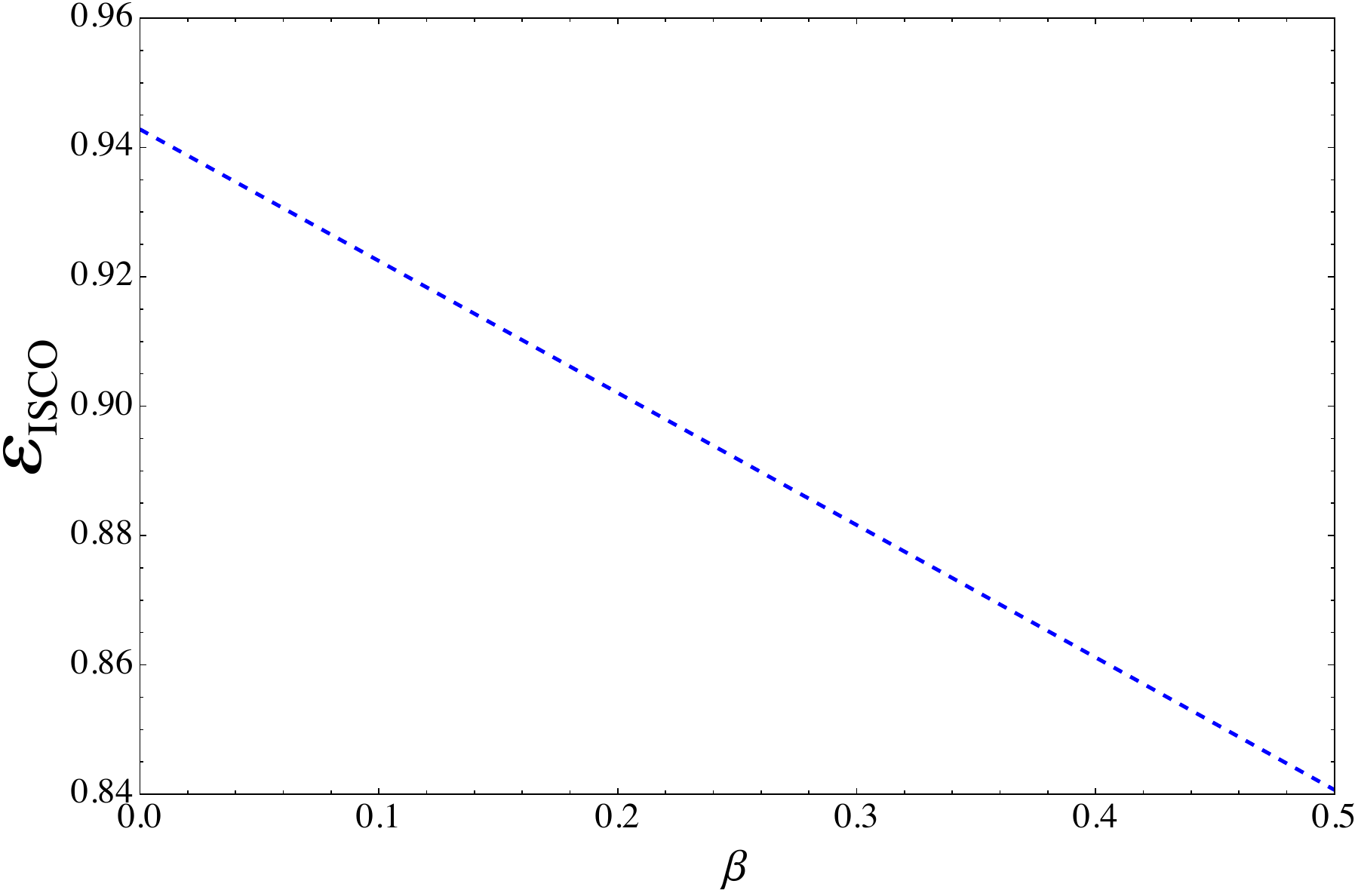}
\caption{
Graphical representation of the ISCO radius $r_{\text{ISCO}}$, ISCO specific angular momentum $l_{\text{ISCO}}$, and the ISCO specific energy $\mathcal{E}_{\text{ISCO}}$ along the magnetic coupling parameter $\beta$. \label{fig.ISCO}}
\end{figure*}

\begin{figure}
\includegraphics[width=0.45\textwidth]{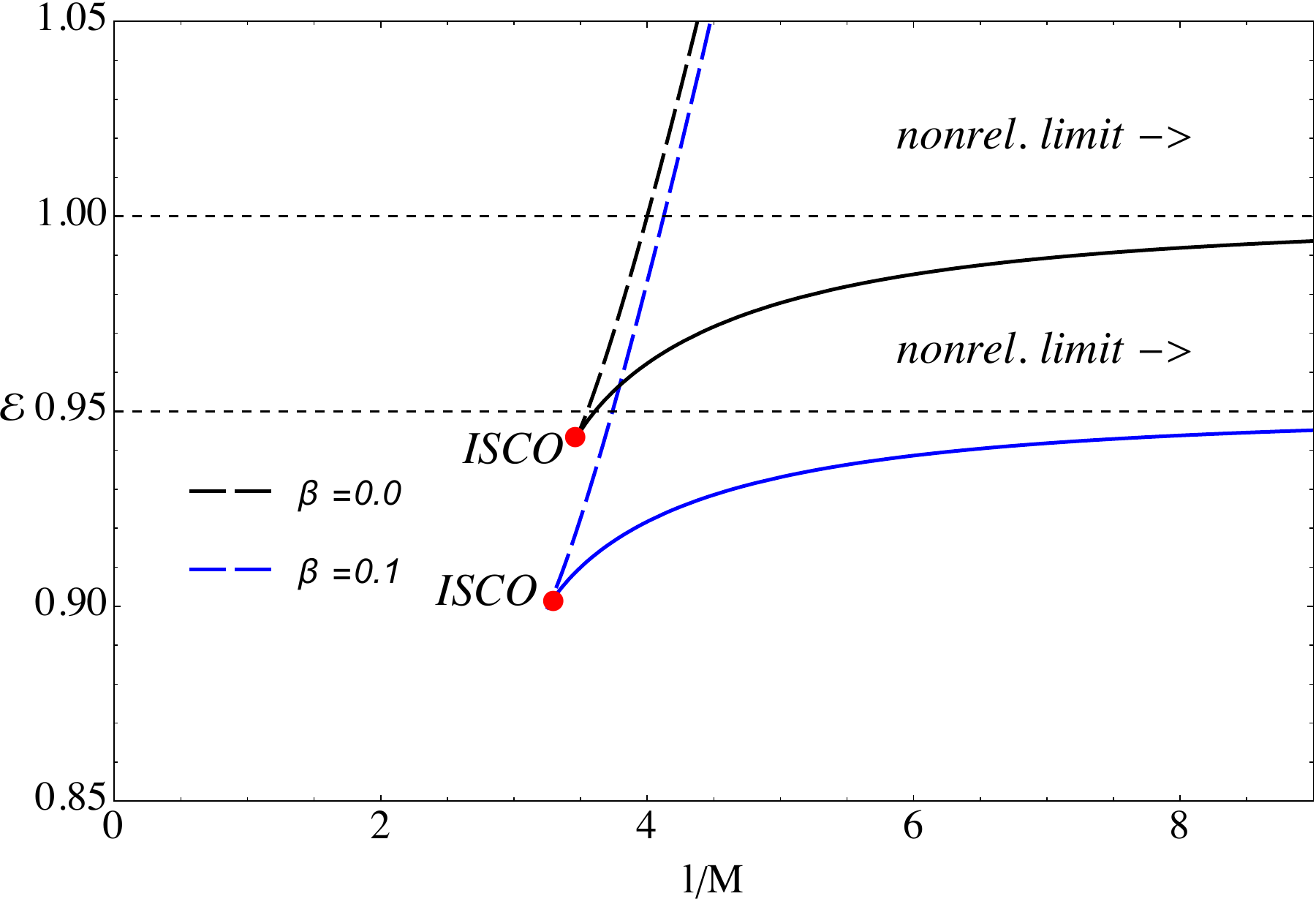}
\caption{Values of the specific energy $\mathcal{E}$ and specific angular momentum $l$ for circular orbits plotted for fixed values of the magnetic coupling parameter $\beta$. Solid lines represent stable circular orbits, while dashed lines represent the radius of unstable circular orbits.}
\label{space}
\end{figure}

Following the same procedure as in Ref. \cite{2024PhRvD.109b4037L} to determine the ($l,\mathcal{E}$) parameter space for circular orbits, in this section, we investigate how this space varies with the magnetic coupling constant $\beta$. Subsequently, by employing the conditions $\mathcal{E}^2=V_{\text{eff}}$ and $V_{\text{eff}}'=0$, we can ascertain the specific energy $\mathcal{E}$ and specific angular momentum $l$ of the magnetized particles as
\begin{eqnarray}\label{12}
l^2&=&\frac{M r_0^2 \left[\beta ^2 f(r_0)+\left(2-3 \beta  \sqrt{f(r_0)}\right)\right]}{2 (r_0-3 M)},
\\ 
\mathcal{E}^2&=&f(r_0)\left(1-\frac{1}{2} \beta  \sqrt{f(r_0)}\right)^2\nonumber\\
&&+\frac{M \left[2-\beta \sqrt{f(r_0)} \left(3 -\beta  \sqrt{f(r_0)}\right)\right]}{2(r_0-3 M)},
\end{eqnarray}
corresponding to circular orbits at the radius $r=r_0$. It is well-known that to identify stable circular orbits, one should examine the second derivative of the effective potential, namely $V_{\mathrm{eff}}''>0$ indicates stable circular orbits. In contrast, $V_{\mathrm{eff}}''<0$ indicates unstable circular orbits. However, our determination of stable and unstable circular orbits depends on $\beta$. Therefore, in Table \ref{Table 1}, we present the values of the critical radii of the unstable circular orbits. It is evident that $V_{\mathrm{eff}}''>0$ for $r_0>r_{\text{max}}$, corresponding to stable orbits, and $V_{\mathrm{eff}}''<0$ for $r_{\text{min}}<r_0<r_{\text{max}}$, corresponding to unstable orbits. Surely, $r_0=r_{\text{min}}$ corresponds to the ISCO. Therefore, from Table \ref{Table 1}, one can easily conclude that increasing the magnetic coupling constant $\beta$ leads to an increase in the value of the ISCO radius. To provide more precise insight, in Fig. \ref{fig.ISCO}, we 
\begin{table}[ht!]
    \centering
    \begin{tabular}{|c|c|c|c|}
     \hline
       $\beta$ & $r_{\min}$ & $r_{\max}$ \\
    \hline
      0.0   & $3M$   & $6M$ \\
      0.1   & $3M$   & $6.09984M$ \\
      0.3   & $3M$   & $6.36414M$ \\
    0.5   & $3M$   & $6.78254M$ \\
   \hline
    \end{tabular}
    \caption{{The radii of stable and unstable circular orbits for different values of the magnetic coupling constant $\beta$.}}
    \label{Table 1}
\end{table}
illustrate how the ISCO radius $r_{\text{ISCO}}$, the specific energy $\mathcal{E}_{\text{ISCO}}$, and the specific angular momentum $l_{\text{ISCO}}$ of the magnetized particle in the ISCO depend on the magnetic coupling constant $\beta$. From Fig. \ref{fig.ISCO}, one can readily conclude that increasing the magnetic coupling parameter $\beta$ results in a decrease in the specific energy $\mathcal{E}_{\text{ISCO}}$ and the specific angular momentum $l_{\text{ISCO}}$ of the magnetized particles in the ISCO. In contrast, increasing the magnetic interaction term $\beta$ leads to a rise in the value of the ISCO radius $r_{\text{ISCO}}$ of the particles with the magnetic dipole moment. Additionally, in Fig. \ref{space}, we have plotted the domain (for the quantities $\mathcal{E}$, $l/M$, and the line representing the non-relativistic limit, i.e. $l/M\to \infty$), within which, magnetized particles can possess circular orbits.
%

\subsection{Analytical solution for the angular trajectories
}

We now endeavor to utilize the classification scheme developed by Levin and Perez-Giz (refer to Ref. \cite{2008PhRvD..77j3005L}) to determine potential values for specific energy $\mathcal{E}$ and specific angular momentum $l$ for a given orbit. To begin this process, we introduce a new variable, $u=1/r$, in Eq. (\ref{11}), which yields
\begin{multline}
\label{13}
    \frac{du}{d\phi} = \pm\frac{\left(1+\beta\sqrt{1-2Mu}\,\right)}{l}\\ 
    \times \sqrt{\mathcal{E}^2-(1-2Mu)\left(1+l^2u^2-\beta\sqrt{1-2Mu}\right)}.
\end{multline}
Subsequently, employing the approximation $\beta\sqrt{1-2Mu}\approx\beta-\beta M u+\beta \mathcal{O}\left(u^2\right)$ and neglecting the $\beta^2u^4$ term, we arrive at the expression
\begin{equation}
  \label{14}
   \left(\frac{du}{d\phi}\right)^2=(1+\beta) \left(au^3-bu^2+cu+d\right),   
\end{equation}
where
\begin{subequations}\label{abcd}
\begin{align}
& a = 2 M \left[(2 \beta +1)-\frac{2 \beta ^2 M^2}{l^2}\right],
\\ 
& b = (\beta +1)+\frac{2 \beta  (1-4 \beta ) M^2}{l^2},
\\
& c = \frac{M \left[2-\beta  \left(5 \beta +2 \mathcal{E}^2-1\right)\right]}{l^2},
\\
& d = \frac{(\beta +1) \left(\beta +\mathcal{E}^2-1\right)}{l^2}.
\end{align}
\end{subequations}
{In the weak interaction limit, Eq. \eqref{14} can be recast as
\begin{equation}
  \label{linerize beta eq.14}
   \left(\frac{du}{d\phi}\right)^2=a_0u^3-b_0u^2+c_0u+d_0,   
\end{equation}
in which
\begin{subequations}\label{abcd}
\begin{align}
& a_0 = 2 M (1+3\beta),
\\ 
& b_0= 1+2\beta\left(1+\frac{M^2}{l^2}\right),
\\
& c_0 =\frac{M}{l^2}\bigl[ 2+\beta(3-2\mathcal{E}^2)\bigr],
\\
& d_0 =\frac{\beta(2\mathcal{E}^2-1)+(\mathcal{E}^2-1)}{l^2}.
\end{align}
\end{subequations}}
From Eq. (\ref{14}), it is evident that for $\mathcal{E}^2<1-\beta$, bound orbits occur, which is the primary focus of this section. We assume $u_1<u_2<u_3$, where $\{u_1,u_2,u_3\}$ represent the real roots of Eq. (\ref{14}). In the scenario where all these roots are positive, in accordance with the right panel of Fig. \ref{effective}, the radii of the particle orbits vary between $r_a={1}/{u_1}$ and $r_p={1}/{u_2}$ (that is, $u_1\leq u\leq u_2$). This situation corresponds to planetary bound orbits (see, for example, Ref. \cite{1998mtbh.book.....C}), wherein the radial position of the test particles oscillates between the apoapsis $u_1$ and the periapsis $u_2$. If one of these roots is negative (for instance, $u_2<0$ and $0<u_1<u_3$), the test particles will either deflect to infinity at the turning point $r_l=1/u_1$ or plunge into the black hole at the turning point $r_s=1/u_3$, as denoted in the right panel of Fig. \ref{effective}. The former type of orbits is termed orbits of the first kind (OFK), while the latter is designated as orbits of the second kind (OSK). These types of orbits are discussed in the next section.

To obtain a general analytical solution for the angular motion, we assume that the magnetized particles commence their motion from $u=u_1$ and $\phi=0$. Then, Eq. (\ref{14}) can be expressed as
\begin{eqnarray}\label{15}
   \phi= \frac{1}{\sqrt{a(1+\beta)}}\int_{u_1}^u\frac{du}{\sqrt{(u-u_1)(u-u_2)(u-u_3)}},
\end{eqnarray}
which yields the solution
\begin{eqnarray}\label{16}
 \phi(u)=\frac{2F\Big(\arcsin{\sqrt{\frac{u-u_1}{u_2-u_1}}},k_0\Big)}{\sqrt{a(1+\beta)(u_3-u_1)}},   
\end{eqnarray}
where 
\begin{equation}
k_0=\sqrt{\frac{u_2-u_1}{u_3-u_1}}.
    \label{eq:k0}
\end{equation}
In Eq. \eqref{16},  $F(x,k)$ is the incomplete elliptic integral of the first kind, with argument $x$ and modulus $k$ \cite{byrd_handbook_1971}. By performing the inversion of the above solution and restoring the change of variable, we arrive at the analytical expression
\begin{eqnarray}\label{17}
    r(\phi)=\frac{1}{u_1+(u_2-u_1)\sn^2(\theta_0, k_0)},
\end{eqnarray}
where $\theta_0\equiv\theta_0(\phi)=\sqrt{a(1+\beta)(u_3-u_1)}\,{\phi}/{2}$. {In the above relation, $\sn(\psi,k)$ denotes the Jacobi elliptic sine function with argument $\psi$ and modulus $k$, and is defined as $\sn(\psi,k)=\sin\xi$, for which
\begin{equation}
    \psi=F(\xi,k)=\int_0^\xi\frac{d t}{\sqrt{1-k^2\sin^2 t}}.
\end{equation}
}

\subsection{Periodic orbits for fixed topological characteristics and the impact of eccentricities 
}

The roots of the characteristic polynomial \eqref{14} can be expressed in the following form \cite{1998mtbh.book.....C}:
\begin{subequations}\label{18}
\begin{align}
    & u_1=\frac{1-e}{\lambda},\\
    & u_2=\frac{1+e}{\lambda},\\
    & u_3=\frac{b}{(1+\beta)a}-\frac{2}{\lambda},
    \end{align}
\end{subequations}
where $e$ and $\lambda$ denote eccentricity and the latus rectum, respectively. For bound orbits, the condition $0\leq e<1$ applies. Then, we can have
\begin{subequations}\label{19}
\begin{align}
   & u_3-u_1=\frac{b}{(1+\beta)a}-\frac{3-e}{\lambda},\\
   & u_2-u_1=\frac{2e}{\lambda}.
\end{align}
\end{subequations}
Accordingly, the exact solution for bound orbits can be written as  
\begin{eqnarray}\label{eq:bound_exact}
    r(\phi)=\frac{\lambda}{1-e+2e \sn^2(\theta_1, k_1)},
\end{eqnarray}
in which
\begin{subequations}\label{eq:theta1_k1}
    \begin{align}
        & \theta_1\equiv\theta_1(\phi)=\sqrt{b-\frac{a(1+\beta)(3-e)}{\lambda}}\,\frac{\phi}{2},\\
        & k_1=\sqrt{\frac{2ea(1+\beta)}{\lambda-a(1+\beta)(3-e)}}.
    \end{align}
\end{subequations}
Considering the oscillation of the radial coordinate between $u_1$ and $u_2$ during one period of planetary orbits, one can determine the associated change in the azimuth angle using Eq. \eqref{16}. This calculation results in,
\begin{eqnarray}
    \label{20}\nonumber
\Delta\phi_u=2\phi(u_2)&=&\frac{4K(k_0)}{\sqrt{a(1+\beta)(u_3-u_1)}}\\
&=&\frac{4K(k_1)}{\sqrt{b-\frac{a(1+\beta)(3-e)}{\lambda}}},
\end{eqnarray}
where $K(k)$ is the complete elliptic integral of the first kind \cite{byrd_handbook_1971}.

Levin and Perez-Giz \cite{2008PhRvD..77j3005L} proposed a taxonomy scheme in the form
\begin{equation}
q+1=\omega+\frac{v}{z}+1=\frac{\Delta\phi_u}{2\pi},
    \label{eq:LevinPerez}
\end{equation}
where $\omega$, $v$, and $z$ represent the whirl, vertex, and zoom numbers, respectively. Subsequently, Equation (\ref{20}) can be reformulated as
\begin{eqnarray}\label{21}
   q+1=\frac{2K(k_1)}{\pi\sqrt{b-\frac{a(1+\beta)(3-e)}{\lambda}}}. 
\end{eqnarray}
Furthermore, for each $e$ and $\lambda$, the corresponding specific angular momentum and specific energy can be found using Eqs. (\ref{14}) and (\ref{18}) as

\begin{eqnarray}
\label{eq:specificE}\nonumber
    \mathcal{E}^2&=&\frac{1}{\lambda  \Psi}\Bigr\{(1+\beta) \Big[2 (1-\beta ) \lambda ^3\\
    &&-4 \beta ^2 \left(1-e^2\right)^2 M^3\Big]+\Theta+\Phi \Bigr\},\\ 
    \label{energy and momentum}
    l^2&=& \frac{(\beta +1) M \Xi }{\Psi },
\end{eqnarray}
{which possess the $\beta$-linearized forms
\begin{eqnarray}
\label{eq:specificE linerized}
    \mathcal{E}^2=\frac{\Theta_0+\Phi_0}{\lambda^2\Psi_0},\quad 
    l^2=\frac{\lambda  M \Xi_0}{\Psi_0},
\end{eqnarray}}
where the coefficients involved are provided in appendix \ref{AppA}.%

\begin{figure*}
\centering
\includegraphics[width=0.45\textwidth]{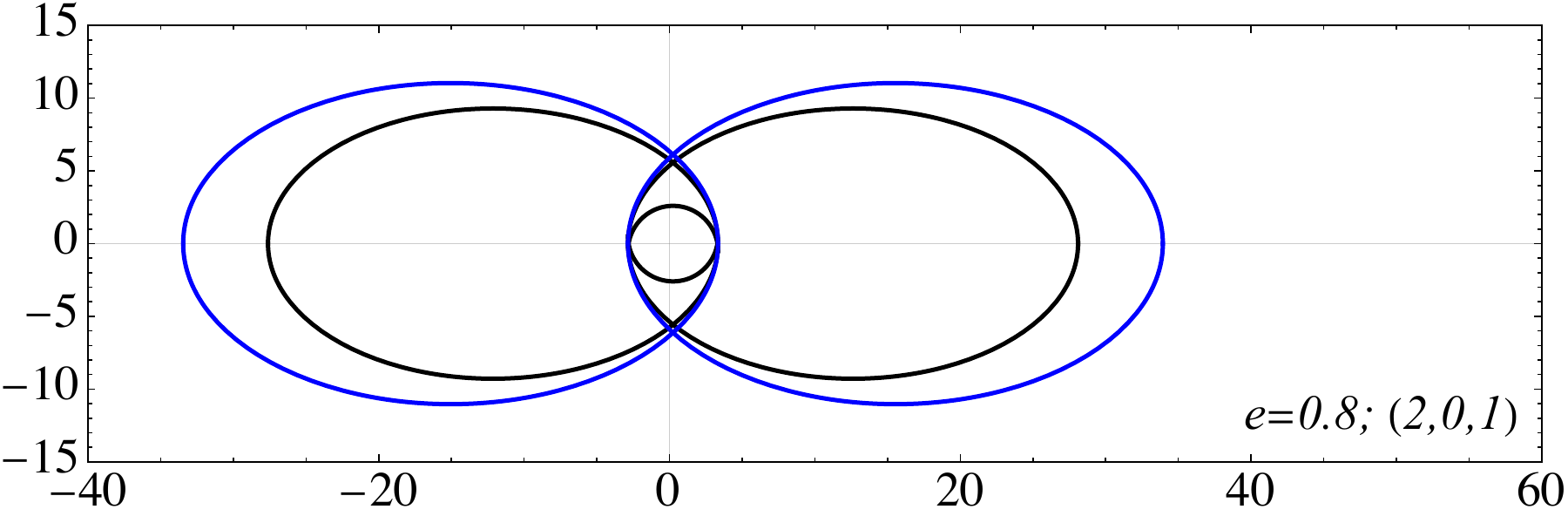}
\includegraphics[width=0.45\textwidth]{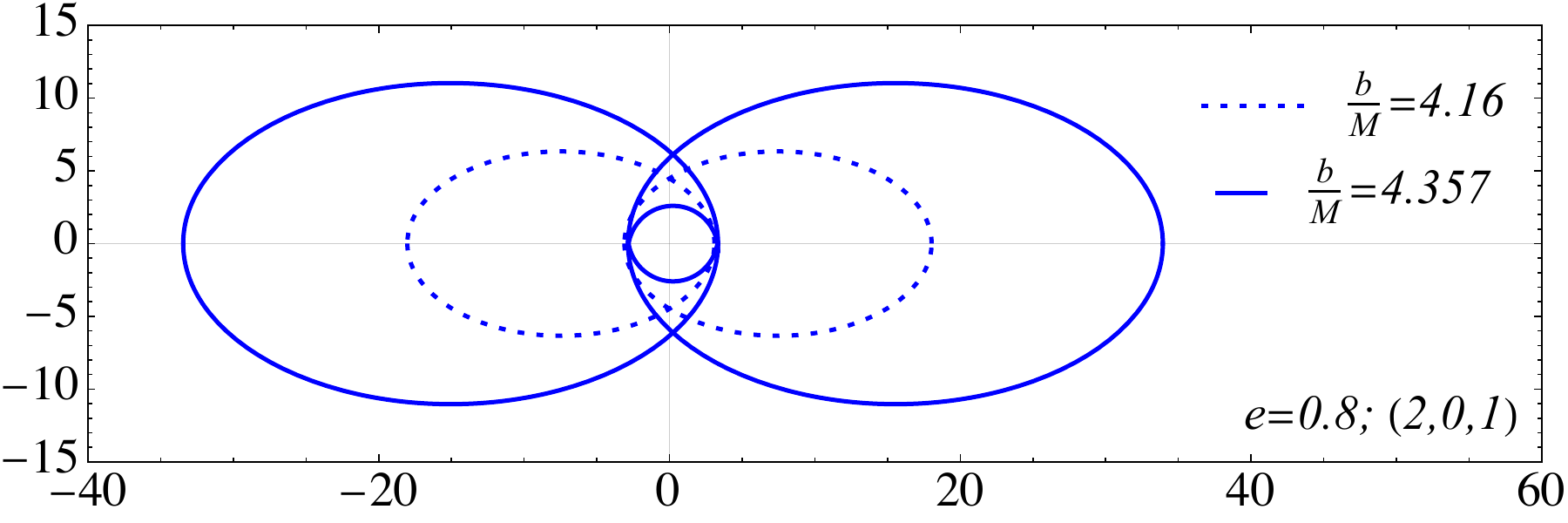}
\includegraphics[width=0.45\textwidth]{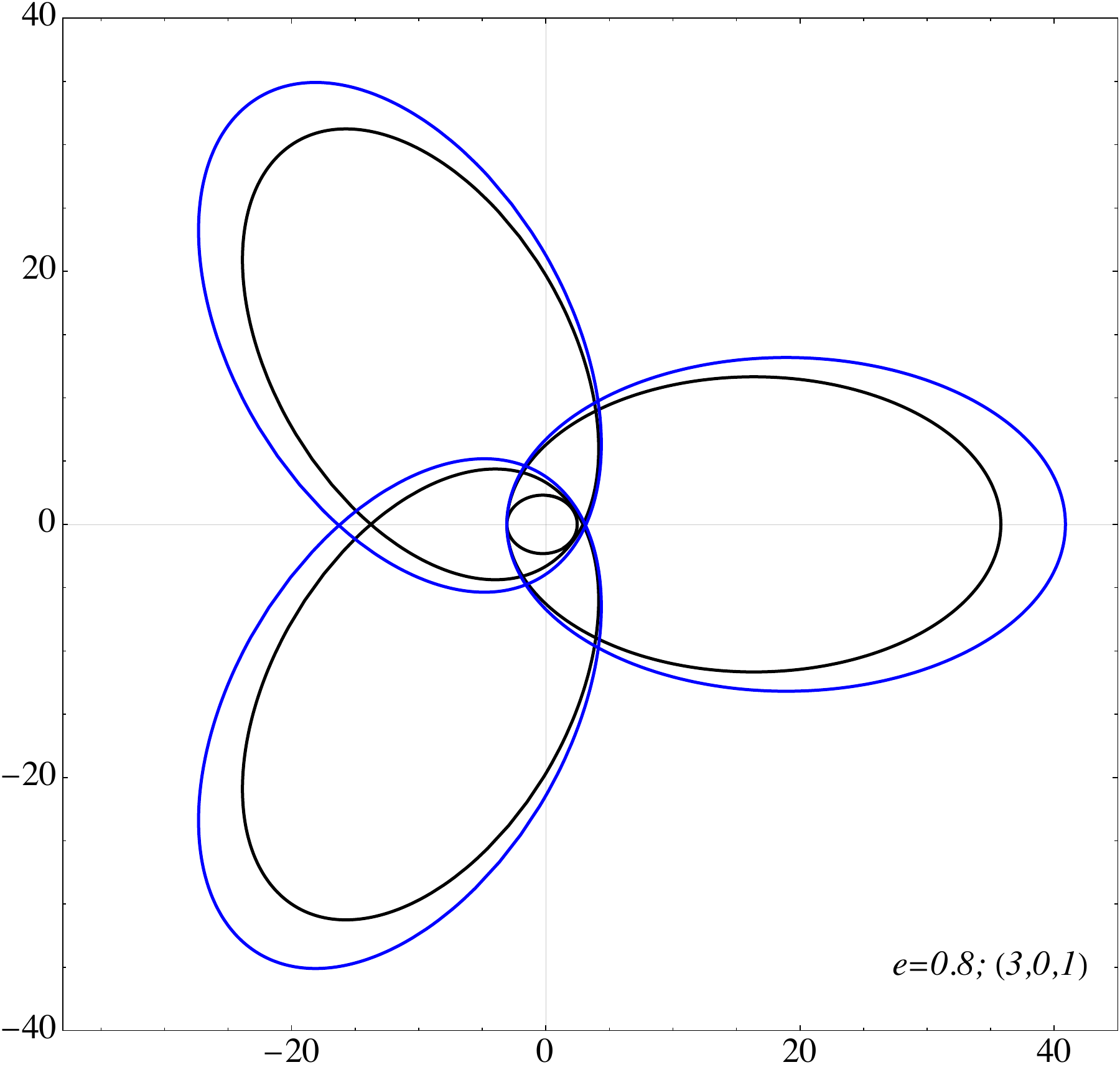}
\includegraphics[width=0.45\textwidth]{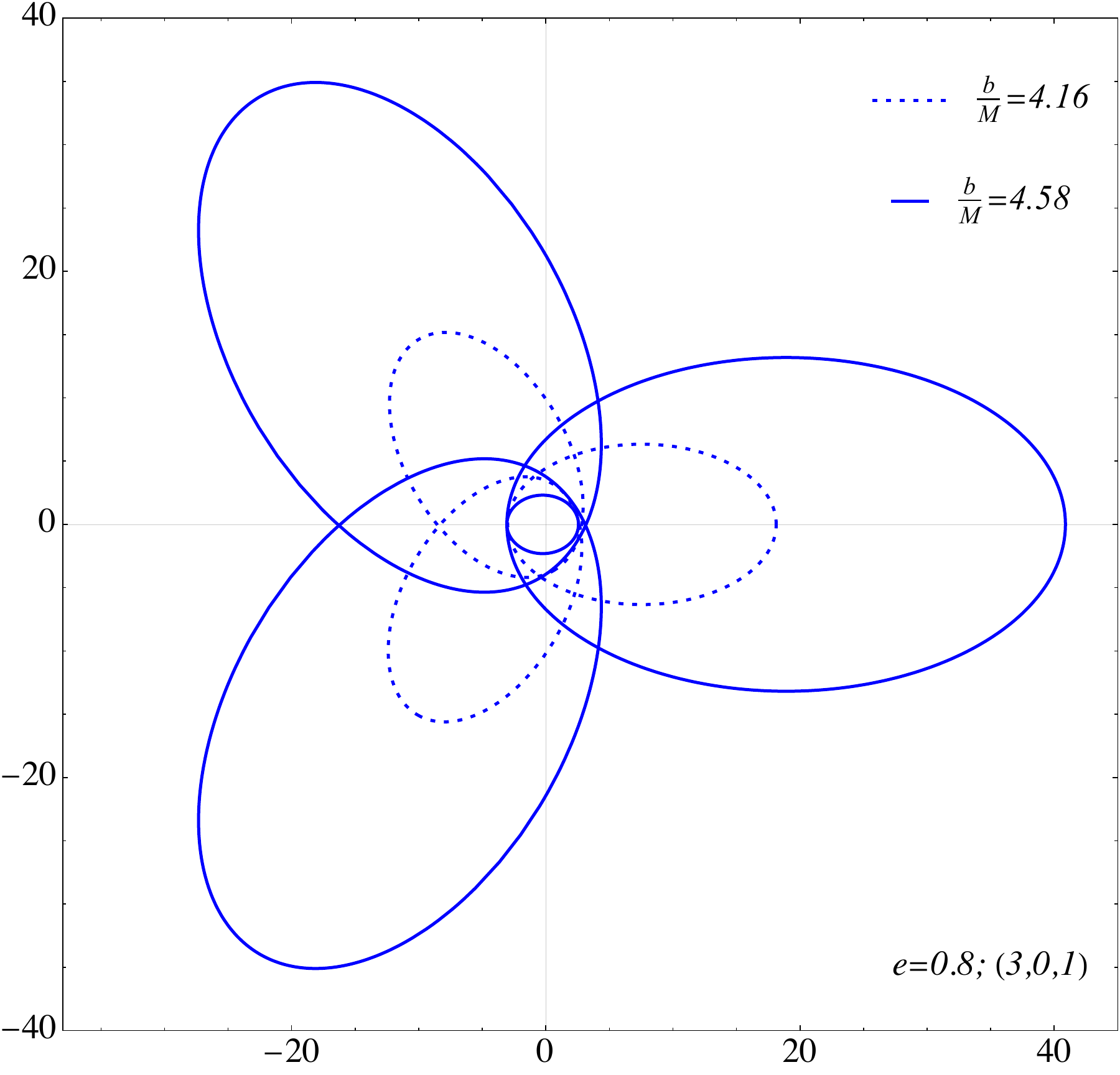}
\includegraphics[width=0.45\textwidth]{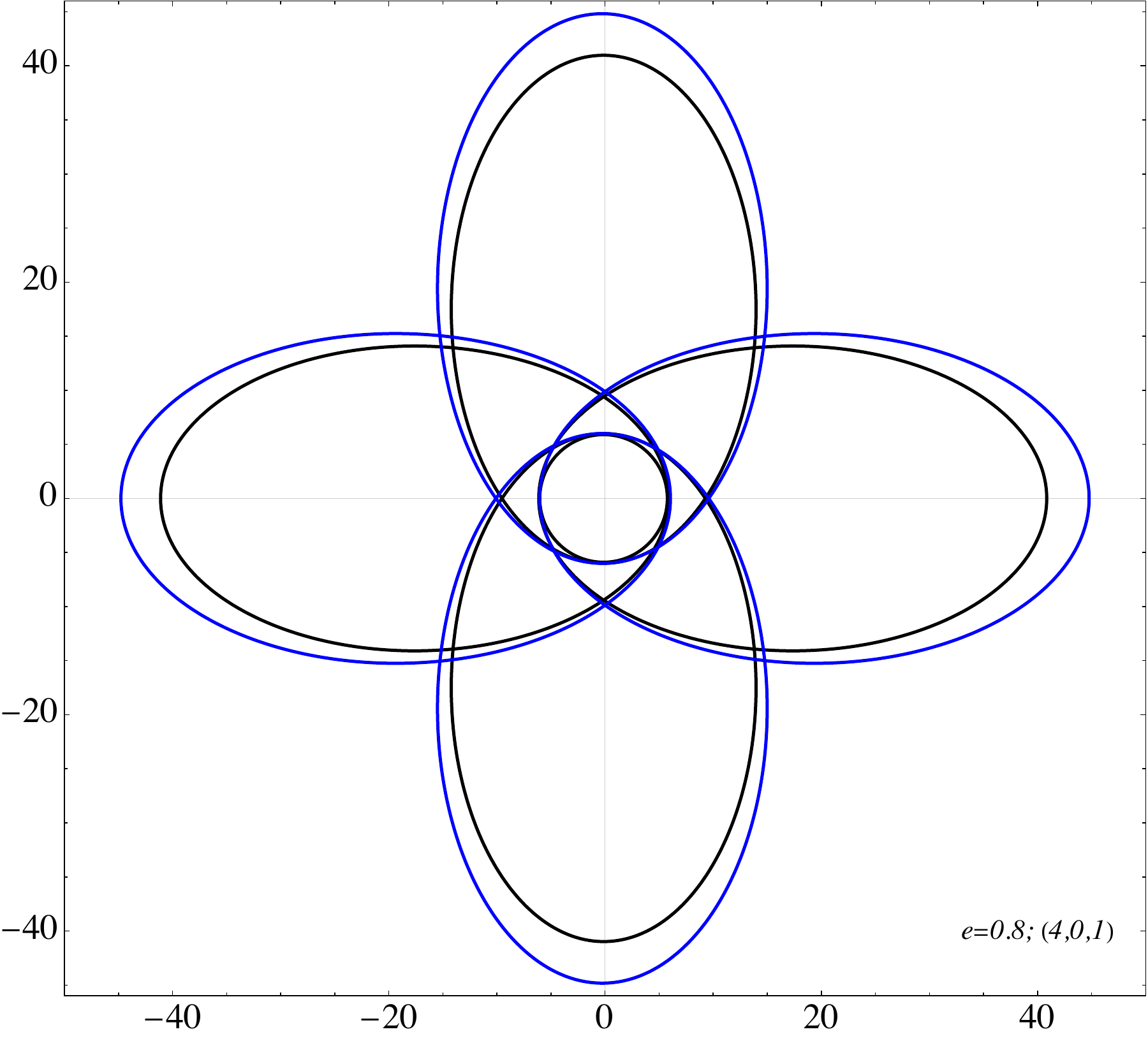}
\includegraphics[width=0.45\textwidth]{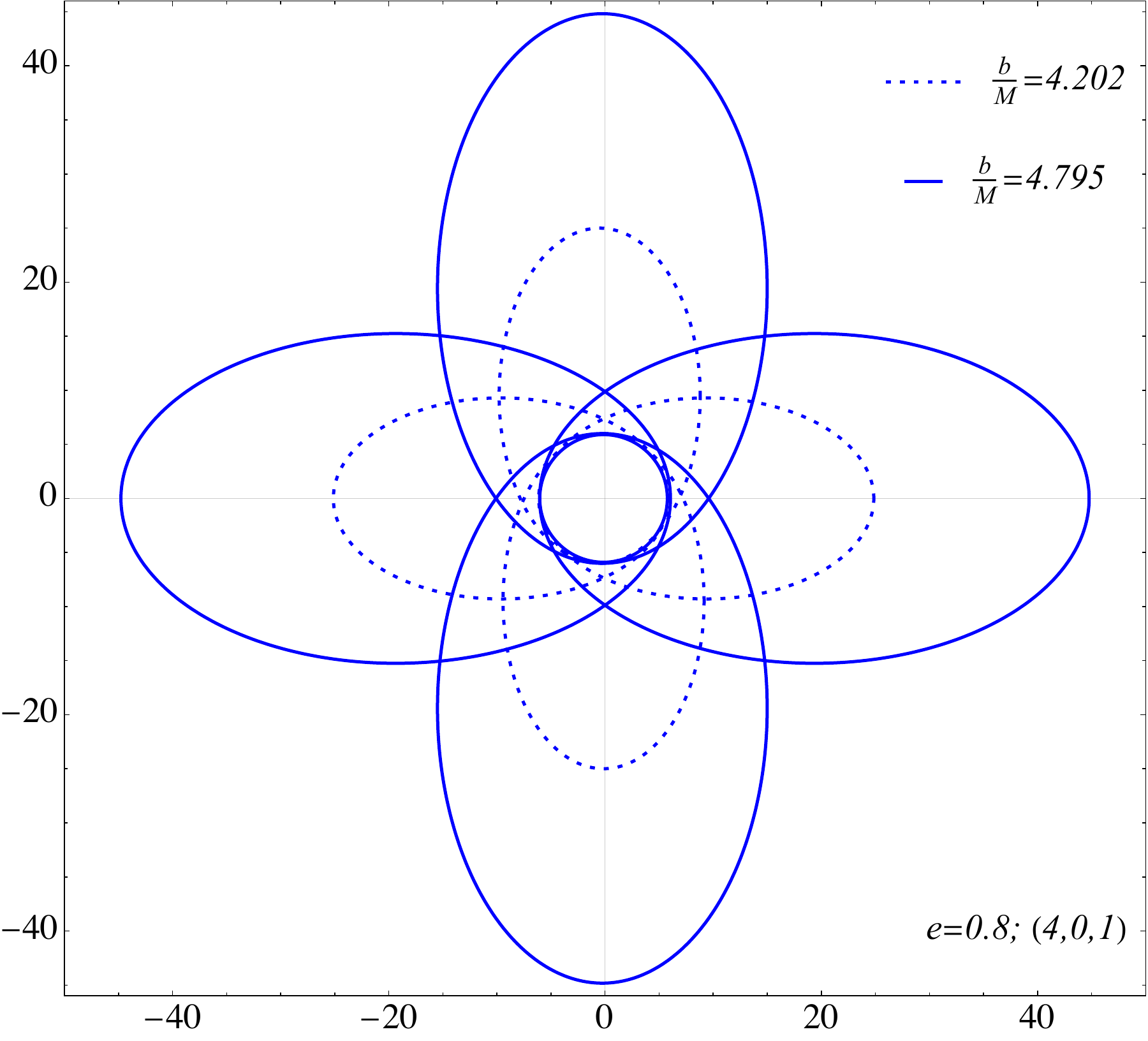}
\caption{{Orbits with fixed eccentricity for different taxonomy schemes. In the left panels, blue orbits correspond to $\beta = 0.1$, while black orbits correspond to $\beta = 0.0$. In the right panels, two different values of the gauge-invariant quantity, the impact parameter $b = \frac{l}{\mathcal{E}}$, are considered. The black hole is assumed to be located at the origin.}}
\label{orbit}
\end{figure*}

Now, we can illustrate the periodic orbits of the magnetized particles for selected values of the periodic orbit parameters $(z, \omega, v)$ and eccentricity $e$. To achieve this, we numerically solve Eq. (\ref{21}) to obtain $\lambda$ and then determine the corresponding specific energy and specific angular momentum using Eq. (\ref{energy and momentum}). In our numerical calculation of $\lambda$, we have disregarded the term $\sim{\beta^2}/{l^2}$ in $a$ and, $b$ as it is considerably smaller than $\sim\beta$. Subsequently, Tables (\ref{Table 2}), (\ref{Table 3}), and (\ref{Table 4}) present numerical values of the latus rectum $\lambda$, along with the corresponding specific energy $\mathcal{E}$ and specific angular momentum $l$, for the periodic orbits $(4,0,1)$, $(3,0,1)$, and $(2,0,1)$, respectively. These tables reveal that the magnetic interaction term $\beta$ leads to an enlargement of the orbits of magnetized particles. This expansion can be attributed to the repulsive interaction between magnetized particles and an external uniform magnetic field $B$. Subsequently, plotting orbits for these selected values is straightforward, as demonstrated in Cartesian coordinates in Fig. \ref{orbit}. The diagrams in this figure illustrate how the magnetic interaction term $\beta$ contributes to the expansion of periodic orbits for test particles with dipole magnetic moment.
\begin{table}[ht!]
    \centering
    \begin{tabular}{|c|c|c|c|c|}
     \hline
       $\beta$ & $e$ & $\lambda$ & $l$ & $\mathcal{E}$ \\
    \hline
      0.0   & 0.5   & $16.7637M$ & $4.56019M$ & 0.978644\\
      0.0   & 0.8   &$16.9124M$ & $4.64227M$ & 0.989591\\ 
      0.1   & 0.5   &$17.8396M$ & $4.39676M$  & 0.930824\\ 
    0.1   & 0.8   & $18.2964M$ & $4.50733M$ & 0.940086\\
    0.1   & 0.8   & $11.1572M$ & $3.92914M$ & 0.935\\
   \hline
    \end{tabular}
    \caption{
The specific energy $\mathcal{E}$ and specific angular momentum $l$ values corresponding to selected eccentricity $e$ and the periodic orbit $(4,0,1)$, with fixed values of the magnetic coupling parameter $\beta$.}
    \label{Table 2}
\end{table}
\begin{table}[ht!]
    \centering
    \begin{tabular}{|c|c|c|c|c|}
     \hline
       $\beta$ & $e$ & $\lambda$ & $l$ & $\mathcal{E}$ \\
    \hline
      0.0   & 0.5   & $13.8243M$ & $4.25126M$ & 0.974472\\
      0.0   & 0.8   &$13.9918M$ & $4.34876M$ & 0.987505\\ 
      0.1   & 0.5   &$15.3592M$ & $4.17356M$  & 0.928205\\ 
    0.1   & 0.8   &$15.8208M$ & $4.29611M$ & 0.938799 \\
    0.1   & 0.8   &$7.60573M$ & $3.86844M$ & 0.93\\
   \hline
    \end{tabular}
    \caption{The same table as in Table \ref{Table 2} for the periodic orbit $(3,0,1)$.}
    \label{Table 3}
\end{table}
\begin{table}[ht!]
    \centering
    \begin{tabular}{|c|c|c|c|c|}
     \hline
       $\beta$ & $e$ & $\lambda$ & $l$ & $\mathcal{E}$ \\
    \hline
      0.0   & 0.5   &$10.9382M$ & $3.94488M$& 0.968567\\
      0.0   & 0.8   &$11.145M$ & $4.06822M$ & 0.984504\\ 
      0.1   & 0.5   &$12.7528M$ & $3.94033M$  & 0.924563 \\ 
    0.1   & 0.8   & $13.2453M$ & $4.0825M$ & 0.936994\\
    0.1   & 0.8   &$7.60573M$ & $3.86844M$ & 0.93\\
   \hline
    \end{tabular}
    \caption{The same table as in Table \ref{Table 2} for the periodic orbit $(2,0,1)$.}
    \label{Table 4}
\end{table}
%

\subsection{Estimating the magnetic coupling parameter $\beta$ using observed data from the S2 star}

Studying the S2 star's orbits provides a unique opportunity to test General Relativity near Sgr A*. It offers insights into stellar dynamics and gravitational forces in the galactic center. The proximity of S2 to the black hole Sgr A* allows exploration of supermassive black hole properties, such as mass and spin. Tracking S2's orbit over time informs our understanding of black hole characteristics and the surrounding environment. Hence, observation data of the motion of the S2 star near Sgr A* offer a pathway to evaluate various physical theories (see, for example, \cite{S21}). In this subsection, we aim to determine the potential values of the magnetic coupling parameter $\beta$ based on the observation data provided in Ref. \cite{s2}. To achieve this, we utilize Eq. (\ref{21}) to compute numerical values of $\beta$. Given numerical values of eccentricity $e$ and latus rectum $\lambda$, we can derive numerical values of $\beta$. For a simple elliptic trajectory, the orbit of the S2 star should be $(1,0,0)$. Thus, Eq. (\ref{21}) can be represented as
\begin{eqnarray}\label{S2 motion}
    \frac{\pi}{2}=\frac{K(k_1)}{\sqrt{b-\frac{a(1+\beta)(3-e)}{\lambda}}}.
\end{eqnarray}
\begin{figure}\centering
\includegraphics[width=0.4\textwidth]{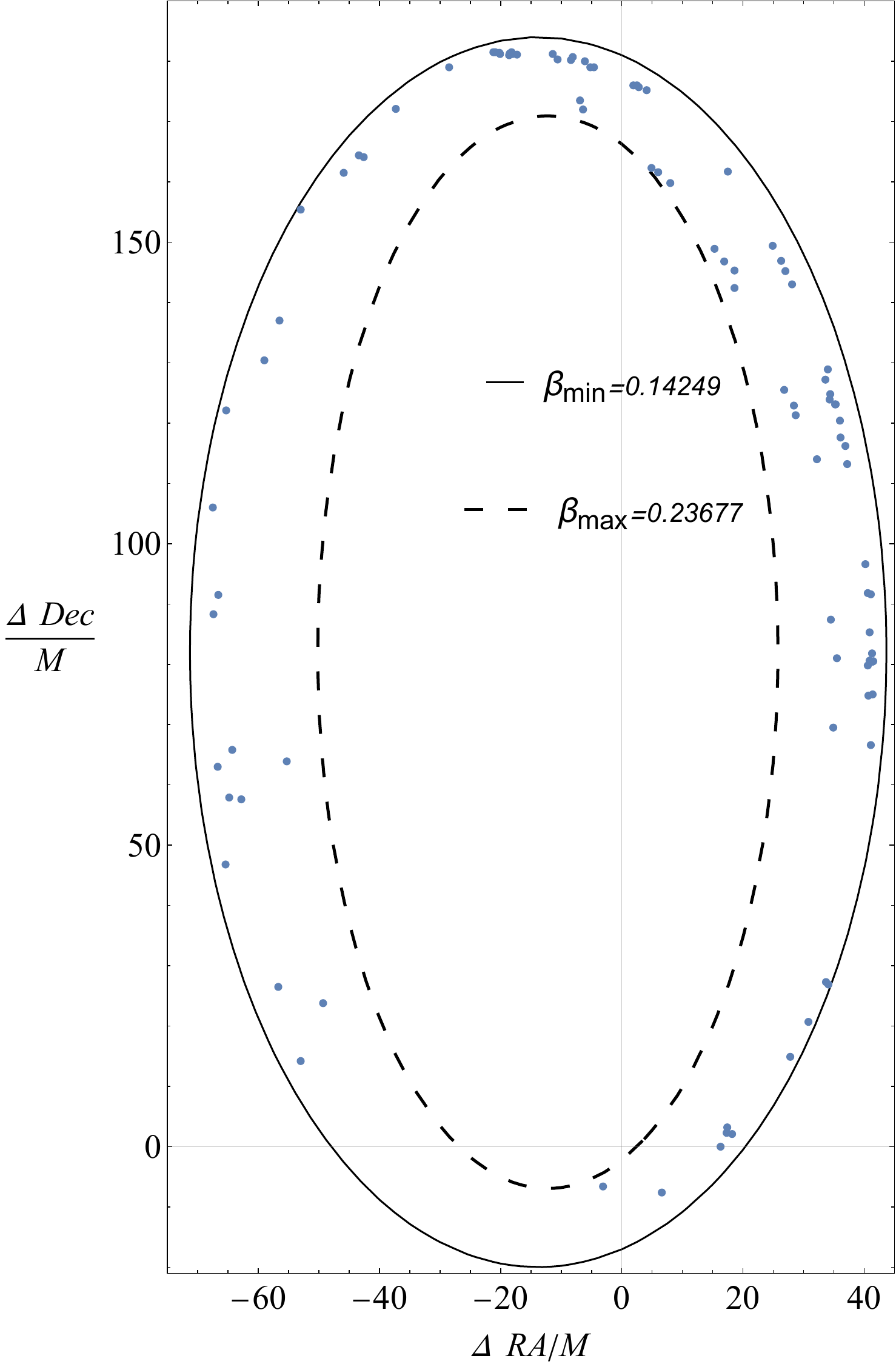}
\caption{Observed trajectory of the star S2 orbiting SgrA*. 
\label{fig. S2}}
\end{figure}
We have depicted the periodic orbit corresponding to the observed trajectory of the S2 star orbiting Sgr A* in Fig. \ref{fig. S2}. In this coordinate system, the $x$ and $y$ axes correspond to offsets in the right ascension ($\Delta \mathrm{RA}/M$) and declination ($\Delta \mathrm{{Dec}}/{M}$) of Sgr A*. Additionally, the origin of the coordinate system coincides with the fitted center of mass of Sgr A*.
\begin{table}[ht!]
    \centering
    \begin{tabular}{|c|c|c|c|c|c|}
     \hline
       $\beta_{\max}$ & $\beta_{\min}$ & $l_{\max}$ & $l_{\min}$ & $\mathcal{E}_{\max}$ & $\mathcal{E}_{\min}$\\
    \hline
      0.419751   & 0.14249   &$5.02989M$ & $7.4095M$ & 0.871573  & 0.923948 \\
   \hline
    \end{tabular}
    \caption{The magnetic coupling parameter $\beta$ values yield periodic orbits for the observed trajectory of the S2 star orbiting Sgr A*, alongside the respective specific energy and specific angular momentum of the S2 star. \label{Table possible}}
\end{table}
Using Eq. (\ref{S2 motion}) together with Eqs. \eqref{eq:specificE} and \eqref{energy and momentum}, we computed the possible values of the magnetic coupling parameter $\beta_{\text{min,max}}$ and, correspondingly, the specific energy $\mathcal{E}_{\text{min,max}}$ and the specific angular momentum $l_{\text{min,max}}$ of the S2 star (see Table \ref{Table possible}). Our calculations indicate that the specific energy and specific angular momentum of the S2 star lie within the ranges $0.871573 \leq \mathcal{E} \leq 0.923948$ and $5.02989M \leq l \leq 7.4095M$, respectively. 

Our findings may help us to determine the limits for the magnetic dipole moment of the S2 star with the mass $\approx 14 M_\odot$ \cite{2017ApJ...847..120H}. The magnetic field in the vicinity of Sgr A* (at the ISCO $6M_{\mathrm{SgrA*}} \simeq 0.24$ AU) is approximately 5-100 G ~\cite{Kunneriath2010AA,Eckart2012AA}, and near the orbit of the magnetar PSR 1745-2900  is about mG~ \cite{Eatough2013Nature}. To estimate the magnetic dipole moment of the magnetized S2 star, we assume the magnetic field near its orbit is about 1 G. The estimated value of the magnetic interaction parameter $\beta=\mu B/m$ is 
\begin{equation}
    \mu\simeq2\beta\times 10^6 \left(\frac{m}{14M_\odot} \right)\left(\frac{B}{1 G} \right)^{-1} {\rm G\cdot cm^3}.
\end{equation}
One may have the possible range of the maximum and minimum of the magnetic dipole using the obtained $\beta_{\max, \min}$ as
\begin{equation}
   \frac{1}{3} \leq \left(\frac{\mu}{10^6 \rm G\cdot cm^3}\right)\leq \frac{1}{2}.
\end{equation}

\section{unbound orbits}\label{sec:unbound}

In this section, we will shift our focus from analyzing bound orbits to exploring unbound trajectories, particularly those involving deflection. These have significant implications for understanding the dynamics of magnetized particles in black hole spacetimes.

\subsection{Unstable circular orbits}

\begin{figure}
\centering
\includegraphics[width=0.425\textwidth]{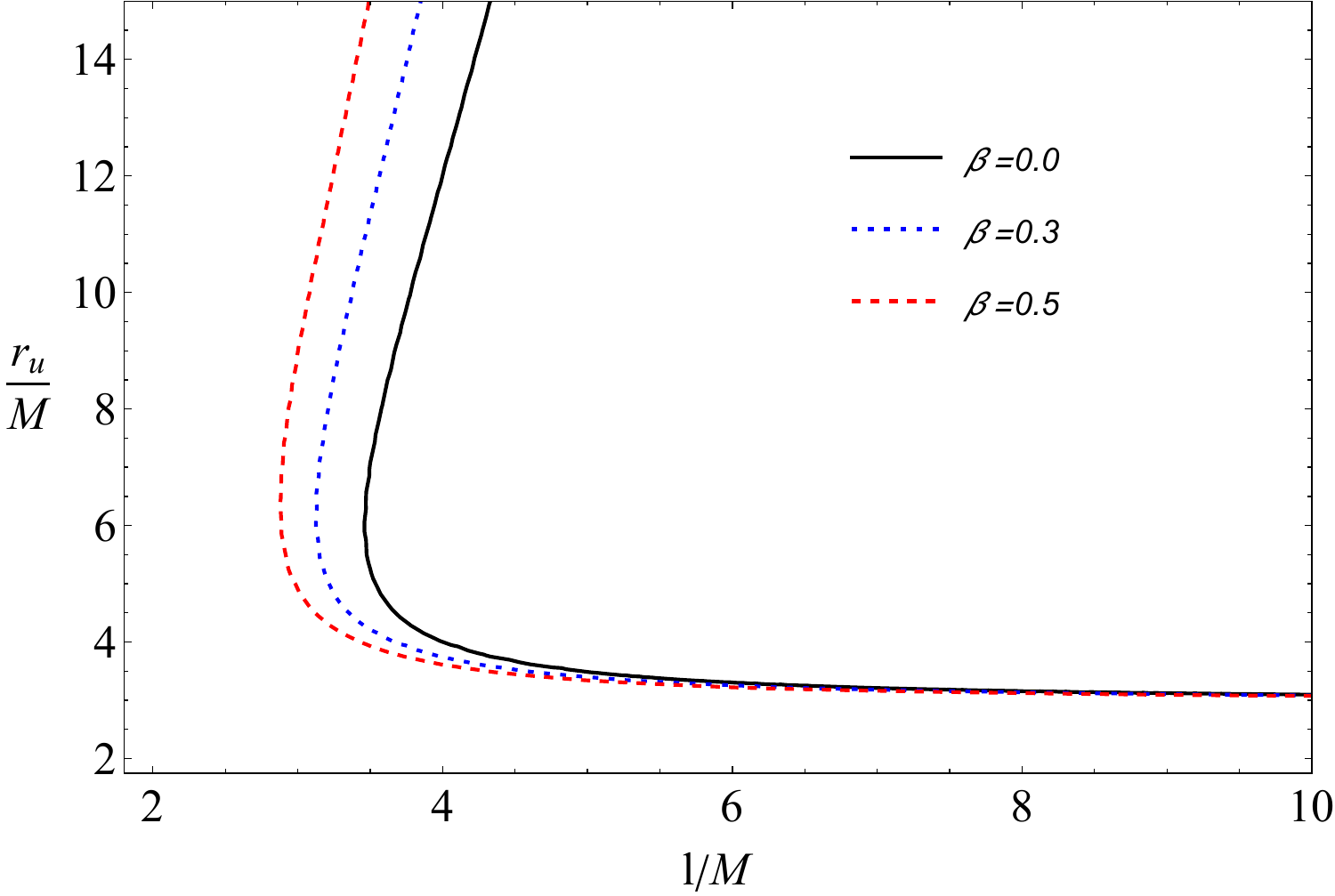}
\caption{The relationship between the radius of unstable circular orbits and specific angular momentum for fixed values of the magnetic coupling constant $\beta$.}
\label{fig:ru}
\end{figure}

According to the right panel of Fig. \ref{effective}, unstable circular orbits occur at the maximum of the effective potential, where the condition $V'_{\mathrm{eff}}=0$ is satisfied. Figure \ref{fig:ru} illustrates the relationship between the radius of unstable circular orbits, $r_u$, and the specific angular momentum, considering fixed values of the coupling parameter $\beta$. As the diagram shows, as the $\beta$ parameter increases, the mutual dependence of the above quantities becomes less noticeable. In other words, for larger values of $\beta$, the changes in $r_u$ depend less strongly on the changes in $l$.
Furthermore, utilizing Eq. (\ref{7}), we can determine the period of unstable circular orbits as measured by both the comoving (proper) and distant observers for a long-term circular orbit. When considering a complete orbit (i.e. $\phi_u=2\pi$), one can obtain the following:
\begin{eqnarray}\label{23}
 T_\tau=\frac{2\pi r_u^2}{l_u}\Big[1+\beta\mathcal{F}(r_u)\Big],\quad
 T_t=\frac{2\pi r_u^2\mathcal{E}_u}{l_uf(r_u)}.
\end{eqnarray}
Furthermore, $l_u$ can be derived using the condition $V'_{\mathrm{eff}}=0$. Furthermore, the expression $\dot{r}=0$ or $\mathcal{E}^2_u=V_{\mathrm{eff}}(r_u)$ leads to the determination of $\mathcal{E}_u$. Therefore, we obtain,
\begin{eqnarray}\label{24}
  && T_\tau = 4\pi r_u\left[1+\sqrt{f(r_u)}\,\beta\right]\Omega(r_u), \\ 
  && T_t = 4\pi r_u\Omega(r_u)\Lambda(r_u), 
\end{eqnarray}
in which 
\begin{subequations}
    \begin{align}
      &  \Omega^2(r) = \frac{r-3 M}{M \left(4-3 \beta  \sqrt{f(r)}\right)},\\   
      & \Lambda^2(r) = \frac{1}{f(r)}\left(\frac{1+l^2}{r^2}-\frac{1}{2} \beta  \sqrt{f(r)} \right).
    \end{align}
\end{subequations}
We continue this section by exploring the deflecting trajectories.

\subsection{OFK}

\begin{figure*}
\includegraphics[width=0.425\textwidth]{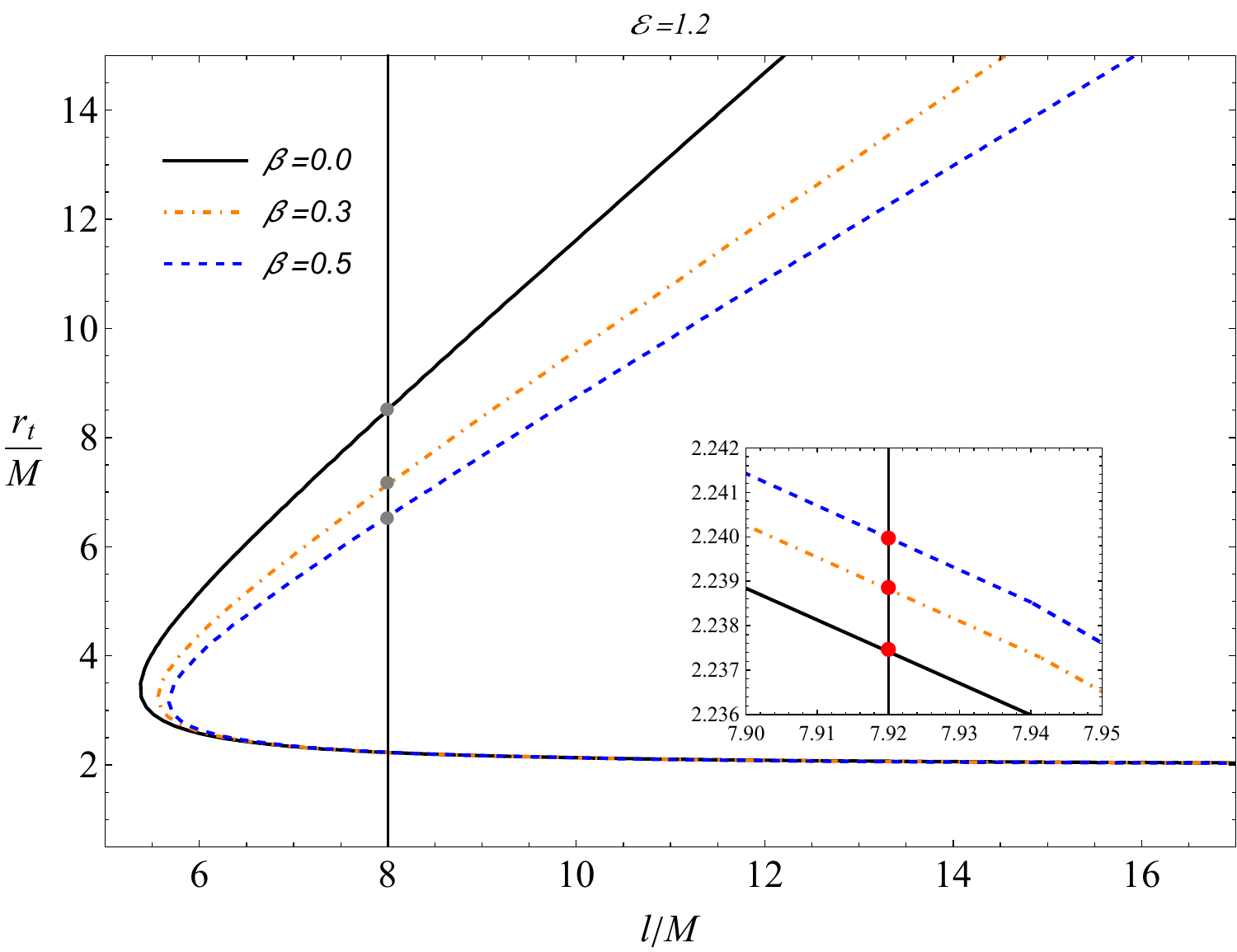}\qquad
\includegraphics[width=0.425\textwidth]{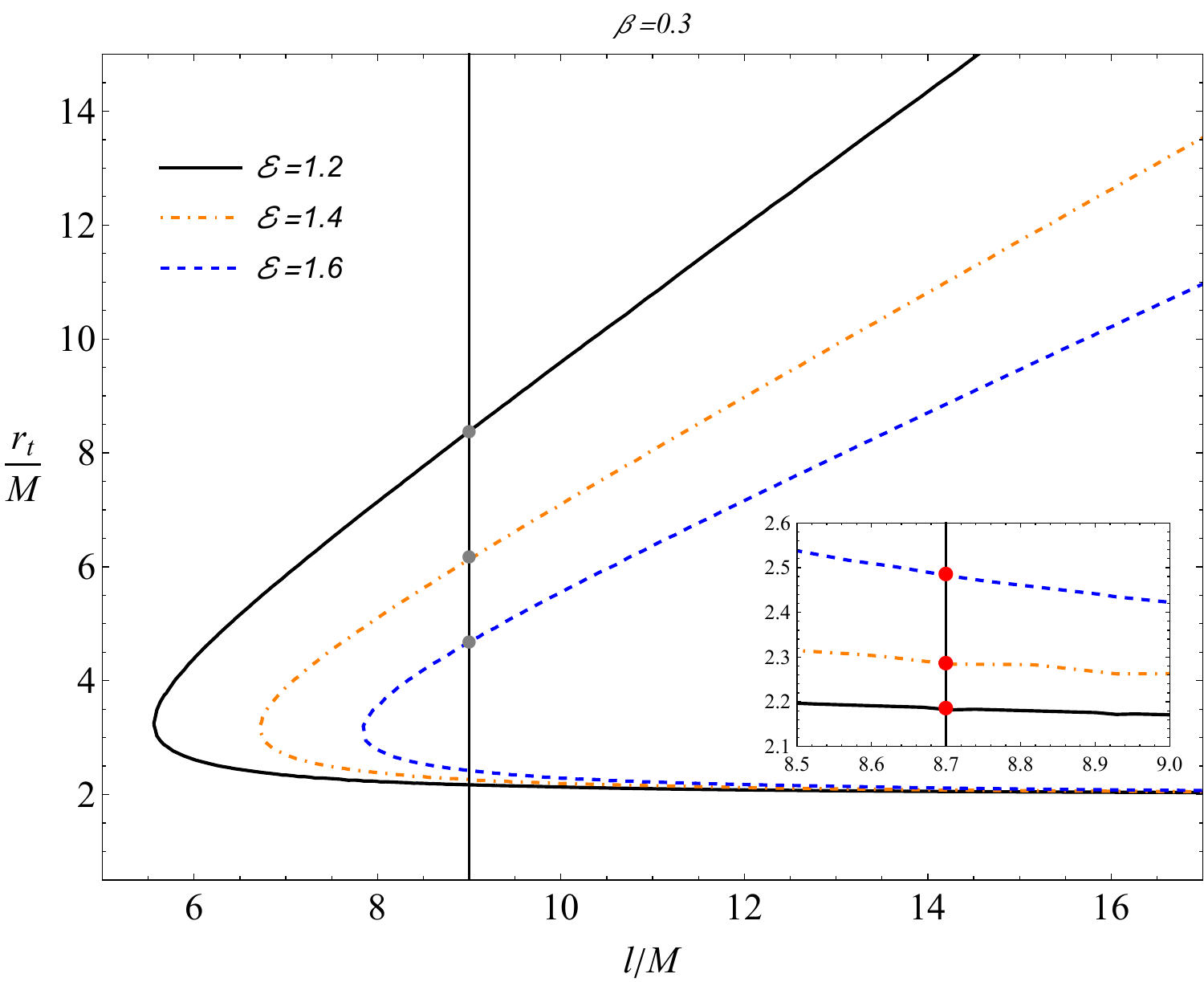}
\caption{The variation of turning point radii with specific angular momentum for fixed values of specific energy $\mathcal{E}$ (left panel), and coupling parameter $\beta$ (right panel). Red points represent the smallest orbital separation $r_s$, while gray points denote the largest orbital separation $r_l$.}
\label{rt}
\end{figure*}

\begin{figure*}
\includegraphics[width=0.425\textwidth]{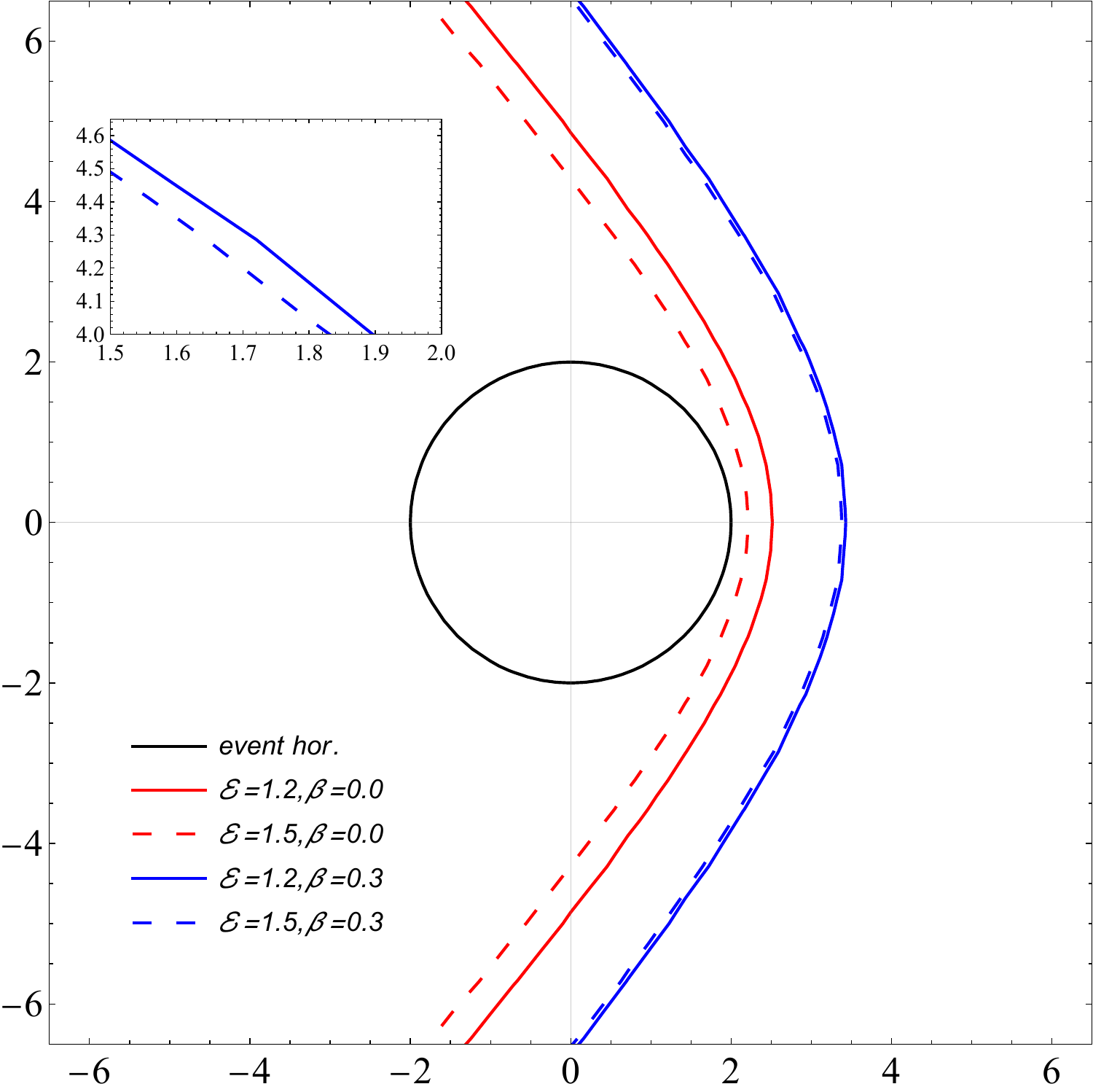}\qquad
\includegraphics[width=0.425\textwidth]{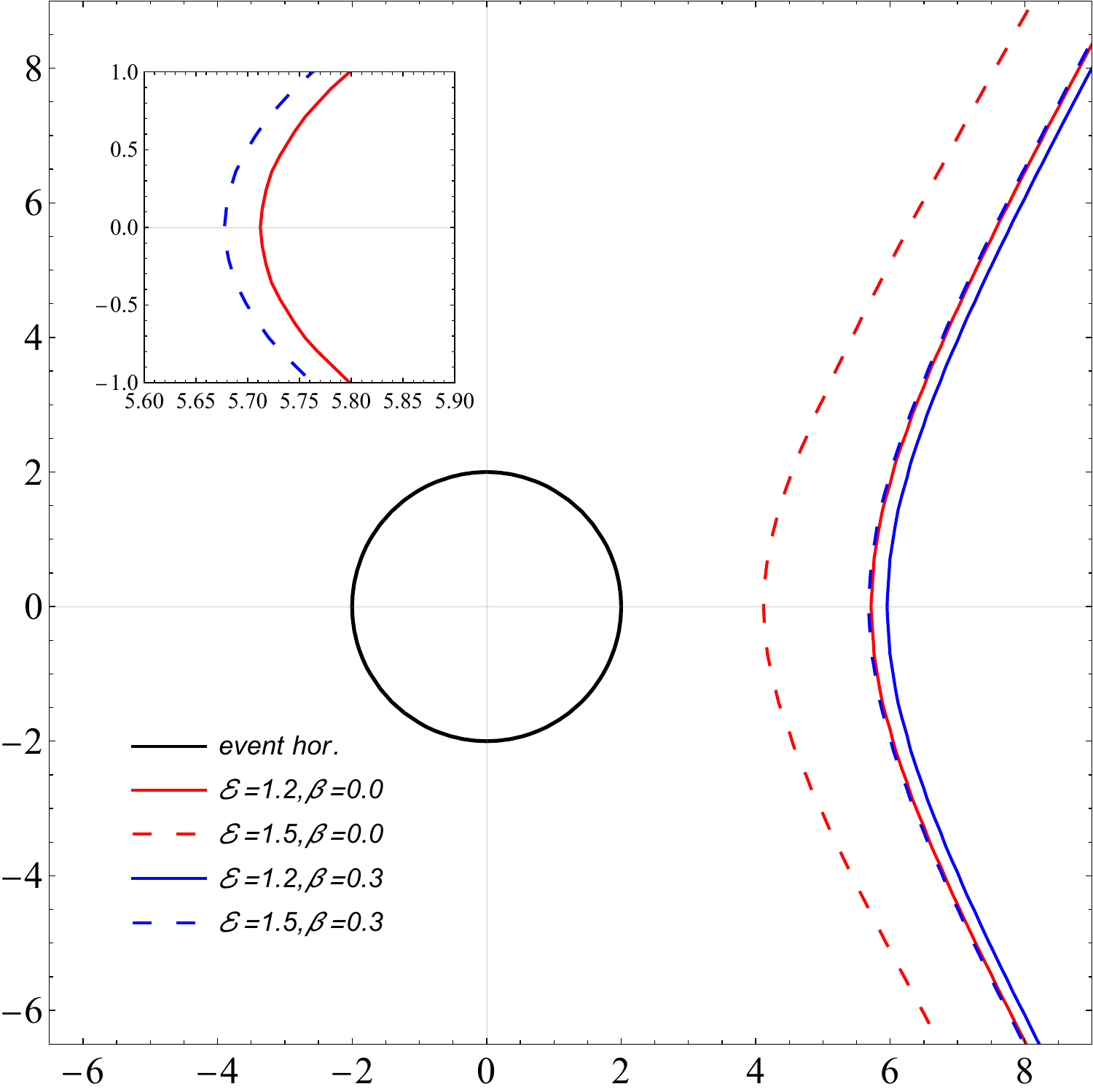}
\caption{The gravitational Rutherford scattering plotted for $e=1.5$ (left panel), and for $e=2$ (right panel).}
\label{hyperbolic}
\end{figure*}

From Eq. (\ref{14}), it is evident that for $\mathcal{E}^2\geq1-\beta$, magnetized particles follow unbound orbits. Utilizing the condition ${dr}/{d\phi}\mid_{r_t}=0$ of a general turning point $r_t$, and Eq. (\ref{11}), we can find the turning points of the particles, which helps to distinguish between OFK and OSK. Subsequently, in Fig. \ref{rt}, we illustrate how the radii of the turning points depend on the angular momentum for fixed values of the specific energy $\mathcal{E}$ and magnetic coupling parameter $\beta$.
These diagrams show that particles exhibit two distinct turning point radii: the smallest orbital separation $r_s$ and the largest orbital separation $r_l$. The latter, $r_l$, corresponds to the OFK. Particles reaching this orbit either proceed to infinity following hyperbolic motion in the $\mathcal{E}^2_u>\mathcal{E}^2\geq1-\beta$ case or engage in periodic orbits when $r_l=r_a$, as observed previously, in the $\mathcal{E}^2<1-\beta$ scenario. Notably, the equation of motion governing OFK is almost identical to that of bound orbits. However, it is crucial to consider that this is one of the roots of Eq. (\ref{14}) is negative for unbound orbits. Consequently, assuming that a magnetized particle initiates its motion from $u=u_2$ and $\phi=0$, Eq. (\ref{16}) can be expressed as
\begin{eqnarray}\label{25}
\phi(u)=\frac{2F\left(\arcsin{\sqrt{\frac{u-u_1}{u_2-u_1}}},k_0\right)-2K\left(k_0\right)}{\sqrt{a(1+\beta)(u_3-u_1)}}.
\end{eqnarray}
Following some straightforward calculations, we obtain the trajectory of the magnetized particles as
\begin{eqnarray}\label{26}
    r(\phi)=\frac{1}{u_1+(u_2-u_1)\sn^2\Big(\theta_0+K(k_0), k_0\Big)}.
\end{eqnarray}
Similarly to Eqs. \eqref{18}, we can express the roots of Eq. (\ref{14}) in the same manner,
however, here $e\geq1$ as we are dealing with hyperbolic orbits. Then Eq. (\ref{26}) can be expressed as
\begin{equation}\label{28}
    r(\phi)=\frac{\lambda}{1-e+2e \sn^2\Big(\theta_1+K( k_1),k_1\Big)}.
\end{equation}
We have determined the corresponding specific energy and angular momentum for each $e$ and $\lambda$ in Eq. (\ref{energy and momentum}). Furthermore, in Table \ref{Table6}, we have provided the values of the latus rectum and the specific angular momentum for the selected values of specific energy $\mathcal{E}$, eccentricity $e$, and magnetic coupling constant $\beta$.
\begin{table}[ht!]
    \centering
    \begin{tabular}{|c|c|c|c|c|}
     \hline
       $\beta$ & $e$ & $\lambda$ & $l$ & $\mathcal{E}$ \\
    \hline
      0.0   & 1.5   &$6.28198M$ & $6.18387M$ & 1.2\\
      0.0   & 1.5   &$5.52617M$ & $10.5156M$ & 1.5\\
      0.0   & 2.0   &$11.4327M$ & $5.43019M$ & 1.2\\
      0.0   & 2.0   &$8.23412M$ & $7.41205M$ & 1.5\\
      0.3   & 1.5   &$8.56076M$ & $13.1119M$  & 1.2\\ 
      0.3   & 1.5   &$8.43699M$ & $19.1212M$  & 1.5\\ 
    0.3   & 2   & $11.9084M$ &$8.70001M$ & 1.2\\
    0.3   & 2   & $11.3611M$ &$12.3606M$ & 1.5\\
   \hline
    \end{tabular}
    \caption{The specific angular momentum and latus rectum values corresponding to selected specific energy and eccentricity, with fixed magnetic coupling parameter.}
    \label{Table6}
\end{table}
Hence, plotting the hyperbolic trajectories of the magnetized particles is now straightforward. In Fig. \ref{hyperbolic}, we depict the OFK for the particles around the black hole. Given the behavior of particles and their deflection from the black hole, the concept of OFK in this context aligns with the magnetic counterpart of the gravitational Rutherford scattering (see Refs. \cite{villanueva_gravitational_2015,fathi_gravitational_2021}). The scattering process can be either repulsive or attractive, depending on the eccentricity $e$ and, correspondingly, the specific energy $\mathcal{E}$ and specific angular momentum $l$ associated with this eccentricity.
The diagrams show that neutral particles approach the black hole more closely than magnetized particles just before heading toward infinity. Similarly to bound orbits, this phenomenon can be attributed to the repulsive nature of the magnetic interaction between magnetized test particles and the external magnetic field $B$. Furthermore, for identical specific energy values, the influence of eccentricity becomes apparent in the transition between attractive and repulsive OFK. An increase in this parameter enhances the tendency of particles to move away from the black hole. 
\raggedbottom

\subsection{OSK}

\begin{figure}[h!]
\includegraphics[width=0.5\textwidth]{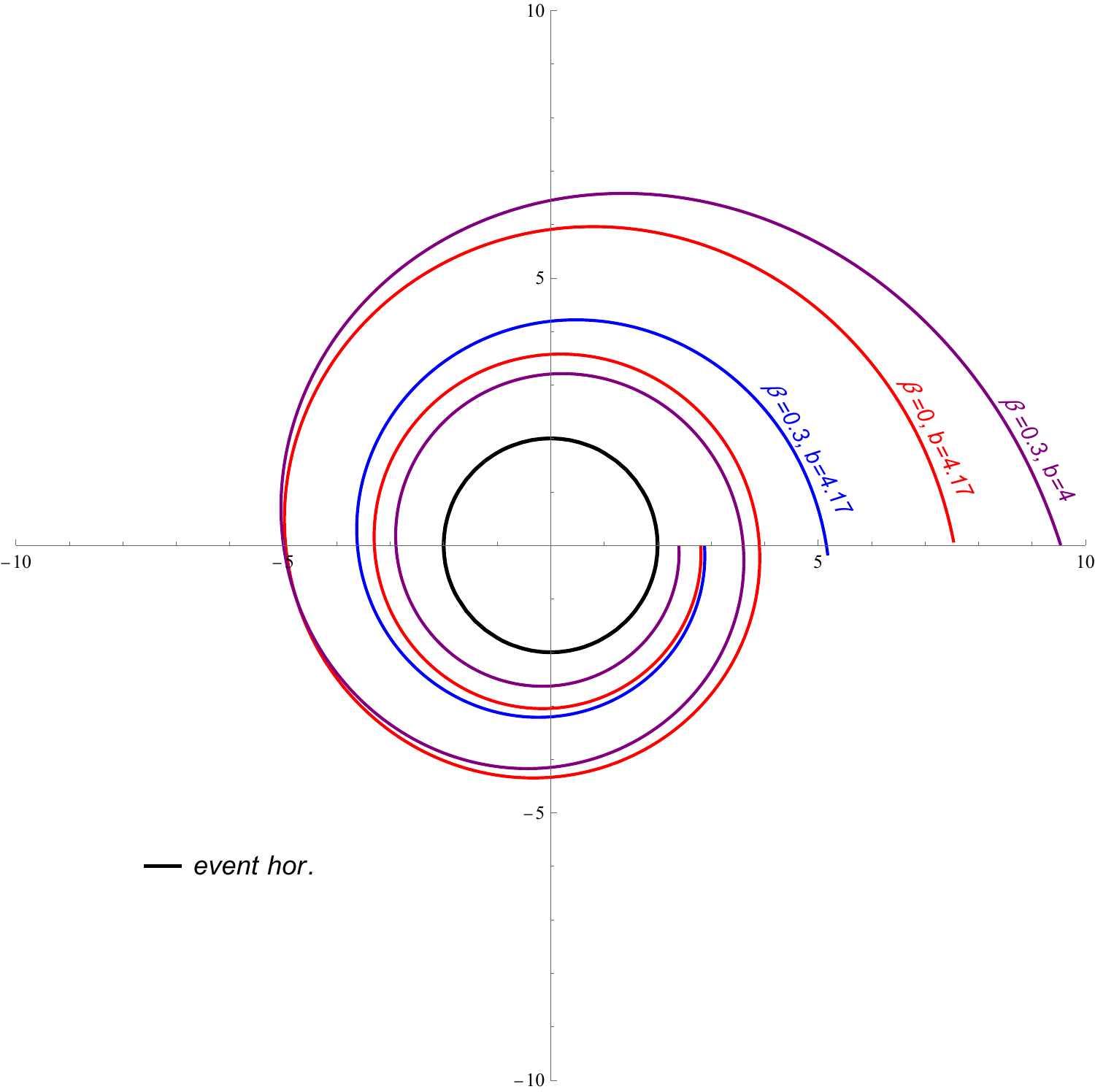}
\caption{The OSK for different values of the $\beta$-parameter and the impact parameter $b$.\label{OSK}}
\end{figure}

As depicted in Fig. \ref{rt}, magnetized particles exhibit two distinct turning points during their unbound orbits. In particular, the smallest orbital separation, denoted as $r_s$, corresponds to the OSK. In this scenario, particles reaching $r_s$ are drawn into the event horizon by spiraling motion. Assuming particles commence their motion from the point $u=u_3={1}/{r_s}$ and $\phi=0$, integration of Eq. (\ref{15}) yields
\begin{multline}\label{29}
\phi(u)=\frac{2}{\sqrt{a(1+\beta)(u_3-u_1)}}\left[F\left(\arcsin{\sqrt{\frac{u-u_1}{u_2-u_1}}},k_0\right)\right.\\
\left.-F\left(\arcsin{\frac{1}{k_0}},k_0\right)\right],
\end{multline}
which enables us to find the analytical solution for the OSK as
\begin{multline}\label{30}
  r(\phi)\\=\frac{1}{u_1+(u_2-u_1)\sn^2\Big(\theta_0+F\left(\arcsin{\frac{1}{k_0}},k_0\right),k_0\Big)}.  
\end{multline}
In Fig. \ref{OSK}, we illustrate the trajectories of magnetized particles approaching the black hole from a distance $r_s$, with fixed values of $\beta$ and the impact parameter $b={l}/{\mathcal{E}}$. From the figure, we infer that a larger impact parameter results in a more noticeable bending of the trajectories as particles approach the black hole, with the turning point positioned at a smaller distance. The magnetic interaction further enhances this bending, especially in the final segment before the particles fall into the black hole. Conversely, smaller impact parameters cause the trajectories to start their approach from a more distant location, indicating a turning point at a greater distance from the black hole.

So far, we have comprehensively analyzed all possible angular trajectories for magnetized test particles and presented fully analytical solutions to the equations of motion. For the completeness of our study, in the next section, we proceed with the analysis of purely radial trajectories.\\

\section{Radial trajectories}\label{sec:radial}

Studying particles in free fall with zero angular momentum offers several advantages for standard general relativistic tests. One key area of investigation is the gravitational clock effect experienced by falling observers in gravitational fields. This forms the theoretical basis for phenomena such as the gravitational redshift-blueshift of light rays passing near a black hole. Another significant aspect is the phenomenon of \textit{frozen} infalling objects, observed by distant observers as they approach the event horizon of a black hole. This stems from the disparities in time measurements between distant observers ($t$) and falling observers ($\tau$), and provides valuable insight into gravitational effects and their implications for our understanding of spacetime dynamics \cite{R1,ryder_2009,zeldovich_stars_2014}. Hence, the radial motion of magnetized particles without angular momentum also presents an interesting subject to explore. 

For purely radial orbits, the effective potential (\ref{10}) adopts the following form:
\begin{eqnarray}\label{Rad.effec}
    V_{\mathrm{eff}}=f(r) \bigr[1-\beta\mathcal{F}(r)\bigr],
\end{eqnarray}
which, together with Eqs. (\ref{8}) and (\ref{10}) enables us to establish relationships for the radial velocities as
\begin{eqnarray}
    \left(\frac{dr}{d\tau}\right)^2&=&\mathcal{E}^2-f(r) \bigr[1-\beta\mathcal{F}(r)\bigr], \label{radial velocity}\\
   \left(\frac{dr}{dt}\right)^2&=&\Bigr\{f(r)\bigr[1+\beta\mathcal{F}(r)\bigr]\Bigr\}^2\nonumber\\
   &&\times\Bigr\{1-\frac{f(r)}{\mathcal{E}^2} \bigr[1+\beta\mathcal{F}(r)\bigr]\Bigr\}.\label{radialvelocity_t} 
\end{eqnarray}
In obtaining analytical representations for the time parameters, we derived power series expansions for Eqs. (\ref{radial velocity}) and \eqref{radialvelocity_t} centered at $\beta=0$ up to the second order. Upon integrating these radial velocities, we can determine the radial dependence of the time parameters for magnetized particles that initiate their motion from the point $r_0$. This procedure yields
\begin{eqnarray}\label{tau}
  &&  \tau(r)=\mathcal{T}_{\tau}(r_0)-\mathcal{T}_{\tau}(r),
\\
  && t(r)=\mathcal{T}_t(r_0)-\mathcal{T}_t(r),
\end{eqnarray}
where
\begin{subequations}\label{eq:TtauTt}
\begin{align}
 & \mathcal{T}_{\tau}(r) = \frac{1}{8} \Big[\mathcal{A}_\tau+\mathcal{B}_\tau-\mathcal{C}_\tau+\mathcal{D}_\tau\Big],  
\\
 & \mathcal{T}_{t}(r) = \frac{1}{8} \mathcal{E}\Big[\mathcal{A}_t+\mathcal{B}_t -\mathcal{C}_t-\mathcal{D}_t\Big].
\end{align}
\end{subequations}
The expansion coefficients are shown in the Appendix \ref{AppB}. 
In Fig. \ref{time}, we depict the dependency of the proper time and coordinate time of magnetized particles on their distance from the black hole. This illustration highlights that while a falling observer may perceive particles crossing the event horizon, the process takes an infinite amount of time for a distant observer.
\begin{figure}[h!]
\centering
\includegraphics[width=0.5\textwidth]{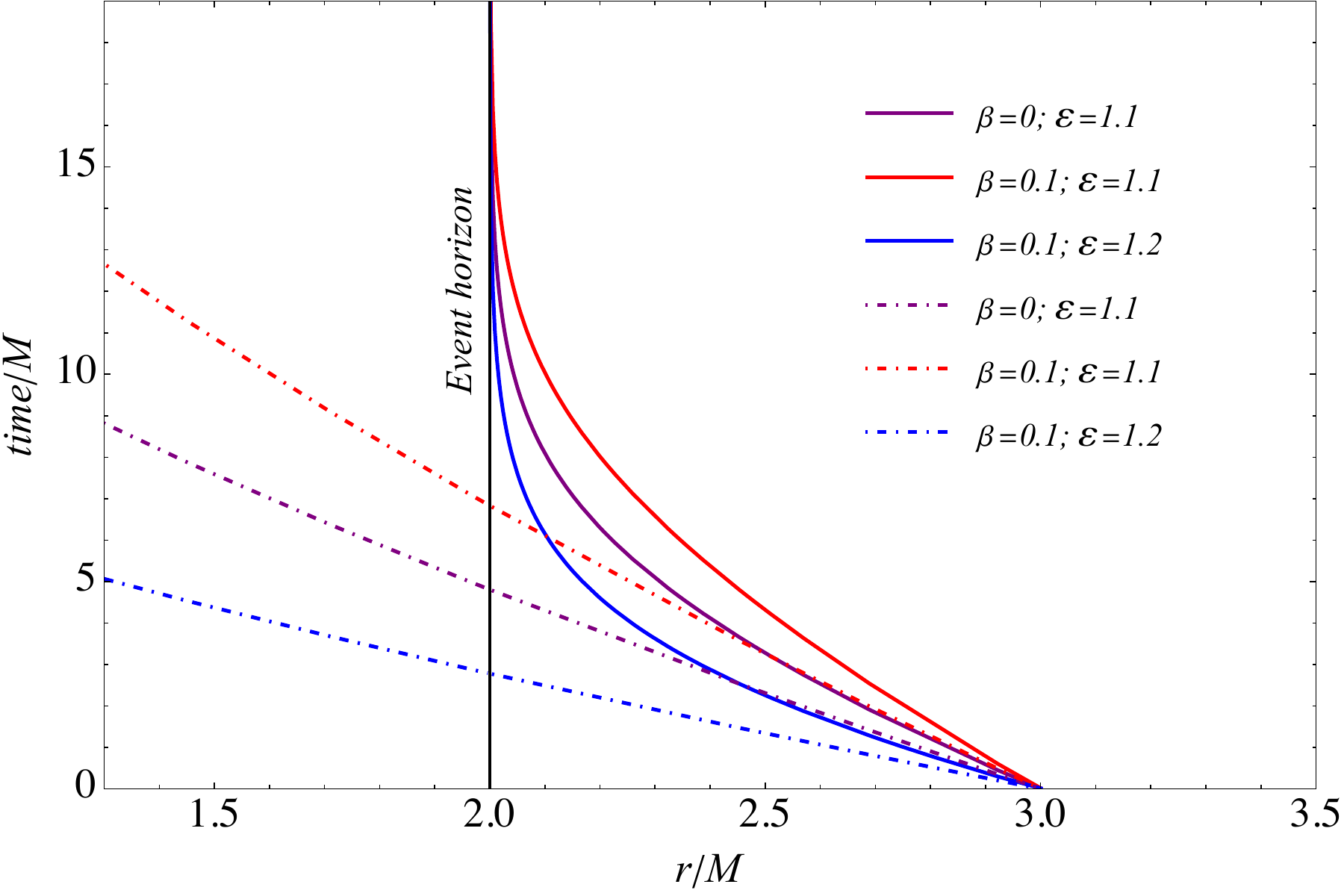}
\caption{Radial profiles of proper time (dot-dashed curves) and coordinate time (solid curves) for test particles on radial orbits, plotted for various values of the $\beta$-parameter and specific energy.}
\label{time}
\end{figure}

Having thoroughly investigated the angular and radial motion of magnetized particles through comprehensive analytical methods, we have explored various aspects of their trajectory dynamics. With this in mind, we close our discussion at this point and proceed to summarize our findings in the next section.

\section{Summary and conclusions}\label{sec:conclusion}

This study investigated particles' bound and unbound orbits with a dipole moment as they approached a magnetized Schwarzschild black hole. Magnetization of the black hole was achieved by assuming proportionality between the four-potential and a space-like Killing vector. We constructed the Lagrangian dynamics for test particles, where the magnetic moment directly contributed to their conjugate momenta, with a magnetic interaction parameter $\beta$ playing a crucial role in determining the magnetic field's effects on particle motion. We derived and analyzed the effective potential experienced by approaching particles, classifying possible bound and unbound orbits based on their specific energy and angular momentum. Bound orbits resembled planetary orbits oscillating between apoapsis and periapsis, with the possibility of complete circularity at the minimum of the effective potential, determining the ISCO. We noted that while the ISCO radius increased with $\beta$, it caused decreases in the specific energy and angular momentum of particles on the ISCO. Analytical solutions to the angular equations of motion for bound orbits were expressed in terms of elliptic integrals and Jacobi elliptic functions, with a precise classification of orbits based on topological characterization. Simulating various bound orbits with different eccentricities revealed topological variations and geometric extensions in the exterior spacetime of the black hole. Comparisons between observed positions of the S2 star around Sgr A* and theoretical predictions were made for two different values of $\beta$, calculating minimum and maximum values for this parameter and specific energy and specific angular momentum of the S2 star. We found that the magnetic dipole moment associated with the S2 star was on the order of $10^6 \,\mathrm{G\cdot cm^{3}}$. Unbound orbits, classified as unstable circular orbits, OFK, and OSK, either deflected from the black hole to infinity or fell inexorably onto the event horizon. The period of unstable circular orbits was calculated for comoving and distant observers, and the OFK was likened to the magnetic counterpart of gravitational Rutherford scattering. For such deflecting trajectories, the characteristic polynomial of the equation of motion had only two positive roots, leading to analytical solutions differing in elliptic integrals' arguments. The scattering of test particles could be attractive or repulsive for the OFK, depending on eccentricity. The OSK trajectory curvature increased with the magnetic interaction parameter, resulting in a more minor turning point. Lastly, we explored purely radial orbits for magnetized particles, obtaining the radial profile of proper time for infalling and coordinating time for distant observers. Studying magnetized particles in black hole spacetimes offers new avenues for investigating relativistic stars at galaxy centers potentially associated with magnetic fields. Expanding models to include black holes with more degrees of freedom could advance research in relativistic astrophysics, a direction for future studies.

\section*{Acknowledgements}
M.F. is supported by Universidad Central de Chile through the project No. PDUCEN20240008. J.R. \& A.A. thank to the Grant F-FA-2021-510 of the Uzbekistan Ministry for Innovative Development.

\appendix

\allowdisplaybreaks


\section{The coefficients of Eqs. \eqref{eq:specificE} and \eqref{energy and momentum}}\label{AppA}

These coefficients are as follows:

\begin{widetext}

\begin{subequations}\label{energy, momentum}
\begin{align}
   & \Phi = 4 \left(\beta +1\right) \left(2 \beta +1\right) \left[\beta  (5 \beta -1)-2\right] \left(e^2+1\right) \lambda  M^2,
\\
   & \Psi  = 8 \beta  \left(\beta +1\right) \left(2 \beta +1\right) \left(1-e^2\right) M^2 - 2 \left(\beta +1\right) \lambda ^2\nonumber\\
   & \qquad\qquad\qquad-2 \lambda  M \Bigr{\{}\beta  \Big[3 \beta  (2 \beta +5)+13\Big]+\Big[\beta  (\beta +1) (2 \beta +3)+1\Big] e^2+3\Bigr{\}},
\\
  & \Xi = \left(-3 \beta ^2-\beta +2\right) \lambda ^3+16 \beta ^3 \left(1-e^2\right) M^3-4 \left(\beta +1\right) \beta ^2 \left(e^2+3\right) \lambda  M^2, 
\\
  &  \Theta = \lambda ^2 M \Bigr\{\beta  \Big[\beta  \Big(6 \beta  (2 \beta +3)-1\Big)+\Big(\beta  \left(4 \beta ^2+6 \beta -7\right)-5\Big) e^2-19\Big]-8\Bigr\},
  \\
  & \Psi_0=2 \left[\lambda -\left(e^2+3\right) M\right]^2,
  \\
  & \Xi_0=(2-3 \beta) \lambda^2+4 M^2 \beta\left(1 +3 e^2\right)+\lambda  M \left[19 \beta +(5 \beta -2) e^2-6\right],
  \\
  & \Theta_0=-2 (\beta -1) \lambda ^4+4 \beta  \left(e^2-5\right) \left(e^2-1\right)^2 M^4+8 \left(e^2-1\right) \lambda  M^3 \left(-6 \beta +e^2+3\right)\nonumber\\
  & \qquad\qquad\qquad+\lambda ^3 M \left[15 \beta +(\beta -2) e^2-14\right],
  \\
  & \Phi_0=\lambda ^2 M^2 \left(\beta  \left(3 e^4+2 e^2-37\right)+32\right).
\end{align}
\end{subequations}

\section{The coefficients of Eqs. \eqref{eq:TtauTt}}\label{AppB}

These coefficients are as follows:

\begin{subequations}\label{eq:B1}
\begin{align}
   & \mathcal{A}_\tau = \frac{r \sqrt{\mathcal{E}^2-f(r)} \left[-4 M^2 \left(3 \beta ^2 \mathcal{E}_1-8\mathcal{E}^2_0\right)-4 M r \mathcal{E}_0 \left(\beta ^2 \mathcal{E}_2-8 \mathcal{E}^2_0\right)+r^2 \mathcal{E}_0 \left(3 \beta ^2+8 \mathcal{E}_0^2\right)\right]}{\mathcal{E}_0^3 \left(2 M+r \mathcal{E}_0\right)^2},
\\ 
   & \mathcal{B}_\tau = \frac{M \left[3 \beta ^2 \left(6 \mathcal{E}^2-1\right)+8 \mathcal{E}_0^2\right] \ln \left(\frac{2 \left(M+r \mathcal{E}_0\right)}{\sqrt{\mathcal{E}_0}}+2 r \sqrt{\mathcal{E}^2-f(r)}\right)}{\mathcal{E}_0^{7/2}},
\\
& \mathcal{C}_\tau = \frac{4 \beta  r \sqrt{f(r)} \sqrt{\mathcal{E}^2-f(r)} \left(M \mathcal{E}_3+r \mathcal{E}_0\right)}{\mathcal{E}_0^2 \left(2 M+r \mathcal{E}_0\right)},
\\
&    \mathcal{D}_\tau = \frac{12 \beta M \mathcal{E}^2\ln \left(2 r \sqrt{f(r)} \sqrt{\mathcal{E}^2-f(r)}-\frac{2 \left[M \left(\mathcal{E}^2-2\right)-r \mathcal{E}_0\right]}{\sqrt{\mathcal{E}_0}}\right)}{\mathcal{E}_0^{5/2}},
\\ 
&    \mathcal{A}_t = \frac{r \sqrt{\mathcal{E}^2-f(r)} \left[4 M^2 \left(\beta^2\mathcal{E}_4+8 \mathcal{E}_0^2\right)+4Mr\mathcal{E}_0 \left(\beta ^2\mathcal{E}_5+8\mathcal{E}_0^2\right)+r^2 \mathcal{E}_0^2 \left(\beta ^2\mathcal{E}_6+8 \mathcal{E}_0^4\right)\right]}{\mathcal{E}_0^3 \left(2 M+r \mathcal{E}_0\right)^2},
\\
&    \mathcal{B}_t = \frac{M \left[\beta ^2 \mathcal{E}_4-8 \mathcal{E}_0^2 \mathcal{E}_7\right] \ln \left(\frac{2 \left(M+r \mathcal{E}_0\right)}{\sqrt{\mathcal{E}_0}}+2 r \sqrt{\mathcal{E}^2-f(r)}\right)}{\mathcal{E}_0^{7/2}}-\frac{4 \beta r\sqrt{f(r)(\mathcal{E}^2-f(r))} \left(M\mathcal{E}_3+r \mathcal{E}_8\right)}{\mathcal{E}_0^2 \left(2 M+r \mathcal{E}_0\right)},
\\
&    \mathcal{C}_t = \frac{4 \beta M\mathcal{E}_9 \ln\left(2 r\sqrt{f(r)(\mathcal{E}^2-f(r))}-\frac{2 \left[M \left(\mathcal{E}^2-2\right)-r\mathcal{E}_0\right]}{\sqrt{\mathcal{E}_0}}\right)}{\mathcal{E}_0^{5/2}},
\\
&    \mathcal{D}_t = \frac{16 M \ln\left(r \left[2\mathcal{E}  \sqrt{\mathcal{E}^2-f(r)}+2 \mathcal{E}_0\right]+2 M\right)}{\mathcal{E}}+\frac{16 M \ln r f(r)}{\mathcal{E}},
\end{align}
\end{subequations}
in which we have defined 
\begin{subequations}\label{eq:B2}
    \begin{align}
        & \mathcal{E}_0=\mathcal{E}^2-1, \\
        & \mathcal{E}_1=2 \mathcal{E}^6-6 \mathcal{E}^4-1, \\
        & \mathcal{E}_2=2\mathcal{E}^6-9 \mathcal{E}^4-3, \\
        & \mathcal{E}_3=4 \mathcal{E}^2+2,\\
        & \mathcal{E}_4=16\mathcal{E}^4-8\mathcal{E} ^2+7,\\
        & \mathcal{E}_5=13 \mathcal{E}^4-10\mathcal{E}^2+7,\\
        & \mathcal{E}_6=8 \mathcal{E}^4-12\mathcal{E}^2+7, \\
        & \mathcal{E}_7=2 \mathcal{E}^2-3,\\
        & \mathcal{E}_8=2\mathcal{E}^4-3r \mathcal{E}^2+r,\\
        & \mathcal{E}_9 = 2 \mathcal{E} ^4-7 \mathcal{E} ^2+2.
    \end{align}
\end{subequations}
\end{widetext}

\bibliography{mybib}    
\end{document}